\documentclass[aps,twocolumn,showpacs,amsmath,amssymb]{revtex4-1}
\usepackage{graphicx}
\usepackage{dcolumn}
\usepackage{bm}
\usepackage{color}

\begin{document}

\title{
Shape evolution of giant resonances in Nd and Sm isotopes
}

\author{Kenichi Yoshida$^{1,2}$ and Takashi Nakatsukasa$^{2}$}
\affiliation{
$^{1}$ Department of Physics, Graduate School of Science and Technology, 
Niigata University, Niigata 950-2181, Japan \\ 
$^{2}$ RIKEN Nishina Center for Accelerator-Based Science, Wako, Saitama 351-0198, Japan
}%

\date{\today}

\begin{abstract}
Giant multipole resonances in Nd and Sm isotopes
are studied by employing the quasiparticle-random-phase approximation 
on the basis of the Skyrme energy-density-functional method. 
Deformation effects on giant resonances are investigated in
these isotopes which manifest a typical nuclear shape change
from spherical to prolate shapes.
The peak energy, the broadening,
and the deformation splitting
of the isoscalar giant monopole (ISGMR) and quadrupole (ISGQR)
resonances agree well with measurements. 
The magnitude of the peak splitting and 
the fraction of the energy-weighted strength in the lower peak of the ISGMR 
reflect the nuclear deformation. 
The experimental data on ISGMR, ISGDR, and ISGQR are consistent with
the nuclear-matter incompressibility $K \simeq  210-230$ MeV 
and the effective mass $m^*_0/m \simeq 0.8-0.9$.
However, the high-energy octupole resonance (HEOR) in $^{144}$Sm seems to
indicates a smaller effective mass, $m^*_0/m \simeq 0.7-0.8$.
A further precise measurement of HEOR is desired to determine the 
effective mass.

\end{abstract}

\maketitle

\section{Introduction}
Giant resonance (GR) is a typical high-frequency collective mode
of excitation in nuclei~\cite{har01}. 
Effects of the nuclear deformation on the GRs have been investigated 
both experimentally and theoretically. 
Among them,
the deformation splitting of the isovector giant dipole resonance (GDR),
due to different frequencies of oscillations along
the major- and minor-axis~\cite{BM2},
is well established. 
A textbook example of the evolution of the GDR 
as a function 
of the mass number can be foundin Refs.~\cite{car71,car74}. 
Emergence of a double-peak structure of 
the photoabsorption cross section of $^{150}$Nd and $^{152}$Sm 
clearly indicates an onset of the deformation in the ground state. 
For the GRs with higher multipolarity,
although the deformation splitting is less pronounced,
the peak broadening has been observed~\cite{har01}. 
The detailed and systematic investigations on the GRs 
would give us a unique information on
the shape phase transition in nuclei.

In contrast to low-energy modes of excitation in nuclei,
the GRs substantially reflect bulk nuclear properties.
Thus, their studies may provide information on the nuclear matter.
The GRs can be qualitatively investigated by using
various macroscopic models,
such as fluiddynamical models which properly take account of deformation
of the Fermi sphere~\cite{RS}.
However, a quantitative description of the GRs requires a microscopic
treatment of nuclear response.
For heavy deformed open-shell nuclei,
the leading theory for this purpose is, currently,
the quasiparticle-random-phase approximation (QRPA) based on
the nuclear energy-density-functional (EDF) method~\cite{ben03}. 
The QRPA based on the deformed ground-state configuration with superfluidity
is able to treat a variety of excitations in the linear regime.
A role of deformation on GRs has been studied by means of 
the deformed QRPA
employing the Gogny interaction in the light mass region~\cite{per08}. 
GRs in heavy systems have been investigated using Skyrme functionals, 
where the separable approximation is employed for the residual interaction~\cite{nes06}, 
and using the relativistic EDF~\cite{pen09}.

The Hartree-Fock-Bogoliubov (HFB) mean
field formulated in the two-dimensional cylindrical coordinates and
the deformed QRPA in the quasiparticle basis
have been developed recently~\cite{yos08}. 
The application, however, was restricted to light systems~\cite{yos09a} 
because of the large computer memory demanded for storing the matrix elements, 
and the time-consuming calculation for 
diagonalizing a non-symmetric matrix of several tens or hundreds of 
thousands of dimensions. 
The deformed Skyrme-QRPA calculation 
utilizing the transformed harmonic oscillator basis 
is also restricted to light nuclei due to the same stumbling block~\cite{los10}. 
Recently, the finite amplitude method~\cite{nak07,avo11}
is applied to the harmonic-oscillator-basis deformed 
QRPA and the calculation for heavy systems becomes possible 
with an inexpensive numerical cost~\cite{sto11}, while it is restricted 
to the $K^{\pi}=0^{+}$ mode so far.

In this article,
we develop a new calculation code of the deformed HFB and QRPA 
for use in the massively parallel computers 
to examine the applicability of the Skyrme-EDF-based QRPA to the excitation modes 
in  heavy deformed systems. 
Using this new parallelized code,
the deformation effects on the GRs in Nd and Sm isotopes will be discussed. 
A part of the results has already appeared in Ref.~\cite{yos11}, 
where we demonstrated that the deformed QRPA can describe well 
the broadening and the deformation splitting of the isovector GDR 
in nuclei undergoing the shape phase transition. 
In the present paper, we perform numerical analysis for the GRs of
multipolarity $L=0-3$ with both isoscalar (IS) and isovector (IV) characters,
and examine the incompressibility and the effective mass
both in spherical and deformed nuclei.
It should be noted that,
in Ref.~\cite{yos10}, the deformation splitting of the giant monopole resonance 
(GMR) in neutron-rich Zr isotopes is predicted by utilizing the calculation
code in this article.

The article is organized as follows: 
In Sec.~\ref{method}, the deformed Skyrme-EDF-QRPA method is recapitulated. 
In Sec.~\ref{para}, some technical details 
to reduce the computational cost are given. 
In Sec.~\ref{result}, results of the numerical analysis of the GRs 
in the Nd and Sm isotopes with shape changes are presented. 
Finally, the summary is given in Sec.~\ref{summary}. 

\section{Deformed HFB + QRPA}\label{method}
\subsection{Basic equations}
The axially deformed HFB in the cylindrical-coordinate space
with the Skyrme EDF and the QRPA in the quasiparticle (qp) representation
can be found in Ref.~\cite{yos08}. 
Here, we briefly describe the outline of the formulation.

To describe the nuclear deformation 
and the pairing correlations, simultaneously, in good account of the continuum,
we solve the HFB equations~\cite{dob84,bul80}
\begin{align}
\begin{pmatrix}
h^{q}(\boldsymbol{r}\sigma)-\lambda^{q} & \tilde{h}^{q}(\boldsymbol{r}\sigma) \\
\tilde{h}^{q}(\boldsymbol{r}\sigma) & -(h^{q}(\boldsymbol{r}\sigma)-\lambda^{q})
\end{pmatrix}
\begin{pmatrix}
\varphi^{q}_{1,\alpha}(\boldsymbol{r}\sigma) \\
\varphi^{q}_{2,\alpha}(\boldsymbol{r}\sigma)
\end{pmatrix} \notag \\
= E_{\alpha}
\begin{pmatrix}
\varphi^{q}_{1,\alpha}(\boldsymbol{r}\sigma) \\
\varphi^{q}_{2,\alpha}(\boldsymbol{r}\sigma)
\end{pmatrix} \label{HFB_equation}
\end{align}
in real space using cylindrical coordinates $\boldsymbol{r}=(\rho,z,\phi)$. 
Here, $q=\nu$ (neutron) or $\pi$ (proton). 
We assume axial and reflection symmetries.
Since we consider the even-even nuclei only, 
the time-reversal symmetry is also assumed. 
A nucleon creation operator $\hat{\psi}^{\dagger}(\boldsymbol{r}\sigma)$ 
at the position $\boldsymbol{r}$ with the intrinsic spin $\sigma$ is  
written in terms of the qp wave functions as
\begin{equation}
\hat{\psi}^{\dagger}(\boldsymbol{r}\sigma)
=\sum_{\alpha}\varphi_{1,\alpha}(\boldsymbol{r}\bar{\sigma})\hat{\beta}^{\dagger}_{\alpha}
+\varphi_{2,\alpha}^{*}(\boldsymbol{r}\sigma)\hat{\beta}_{\alpha}.
\end{equation}
The notation $\varphi(\boldsymbol{r}\bar{\sigma})$ is defined by 
$\varphi(\boldsymbol{r}\bar{\sigma})=-2\sigma \varphi(\boldsymbol{r}-\sigma)$. 

For the mean-field Hamiltonian $h$, we mainly employ the SkM* functional~\cite{bar82}. 
For the pairing energy, we adopt the one in Ref.~\cite{yam09}
that depends on both
the isoscalar ($\varrho$) and the isovector ($\varrho_{1}$) densities, 
in addition to the pairing density ($\tilde{\varrho}$):
\begin{equation}
\mathcal{H}_{\mathrm{pair}}(\boldsymbol{r})
=\dfrac{V_{0}}{4}\sum_{q}\mathrm{g}_{q}[\varrho,\varrho_{1}]
[\tilde{\varrho}(\boldsymbol{r})]^{2}
, \label{pair_int}
\end{equation}
with
\begin{equation}
\mathrm{g}_{q}[\varrho,\varrho_{1}]=1-\eta_{0}\dfrac{\varrho(\boldsymbol{r})}{\varrho_{0}}
-\eta_{1}\dfrac{\tau_{3}\varrho_{1}(\boldsymbol{r})}{\varrho_{0}}
-\eta_{2}\left[\dfrac{\varrho_{1}(\boldsymbol{r})}{\varrho_{0}}\right]^{2}.
\end{equation}
Here $\varrho_{0}=0.16$ fm$^{-3}$ 
is the saturation density of symmetric nuclear matter, 
with the parameters ($\eta_{0}, \eta_{1}$ and $\eta_{2}$) given in 
Table~III of Ref.~\cite{yam09}.
Because of the assumption of the axially symmetric potential,
the $z-$component of the qp angular momentum, $\Omega$,
is a good quantum number. 
Assuming time-reversal symmetry and reflection symmetry
with respect to the $x-y$ plane,
the space for the calculation can be reduced into the one with
positive $\Omega$ and positive $z$ only. 

Using the qp basis obtained
as a self-consistent solution of the HFB equations (\ref{HFB_equation}),
we solve the QRPA equation in the matrix formulation~\cite{row70}
\begin{equation}
\sum_{\gamma \delta}
\begin{pmatrix}
A_{\alpha \beta \gamma \delta} & B_{\alpha \beta \gamma \delta} \\
-B_{\alpha \beta \gamma \delta} & -A_{\alpha \beta \gamma \delta}
\end{pmatrix}
\begin{pmatrix}
X_{\gamma \delta}^{i} \\ Y_{\gamma \delta}^{i}
\end{pmatrix}
=\hbar \omega_{i}
\begin{pmatrix}
X_{\alpha \beta}^{i} \\ Y_{\alpha \beta}^{i}
\end{pmatrix} \label{eq:AB1}.
\end{equation}
The residual interaction in the particle-hole (p-h) channel appearing
in the QRPA matrices $A$ and $B$ is
derived from the Skyrme EDF. 
The residual Coulomb interaction is neglected because of the computational limitation.
We expect that the residual Coulomb plays only a minor role~\cite{ter05,sil06,eba10,nak11}. 
In Ref.~\cite{sil06}, 
effects of neglecting the residual Coulomb interaction are
discussed in details:
The centroid energy of the GDR can be shifted by about 400 keV at maximum.
However, this amount of change does not affect the discussion 
in the present paper.
We also drop the so-called $``{J}^{2}"$ term $C_{t}^{T}$ both in 
the HFB and QRPA calculations. 
The residual interaction in the
particle-particle (p-p) channel is derived from the pairing EDF~(\ref{pair_int}). 
It is noted here that we have an additional contribution to the 
residual interaction in the p-h channel coming from the pairing EDF~(\ref{pair_int}) 
because of the squared $\eta_2$ term in Eq. (\ref{pair_int}) 
(see Appendix A). 

\subsection{Details of the numerical calculation}\label{para}
For solution of the HFB equations (\ref{HFB_equation}), 
we use a lattice mesh size $\Delta\rho=\Delta z=0.6$ fm and a box
boundary condition at $\rho_{\mathrm{max}}=14.7$ fm, $z_{\mathrm{max}}=14.4$ fm. 
The differential operators are represented by use of the 11-point formula of finite difference method. 
Since the parity ($\pi$) and  the magnetic quantum number ($\Omega$) 
are good quantum numbers, 
the HFB Hamiltonian 
becomes in a block diagonal form with respect to each $(\Omega^{\pi},q)$ sector.
The HFB equations for each sector are solved independently with 48 processors for 
the qp states up to $\Omega=23/2$ with positive and negative parities. 
Then, the densities and the HFB Hamiltonian are updated, 
which requires communication among the 48 processors.
The modified Broyden's method~\cite{bar08} is utilized to calculate new densities.
The qp states are truncated according to the qp
energy cutoff at $E_\alpha \leq 60$ MeV. 

We introduce the additional truncation for the QRPA calculation,
in terms of the two-quasiparticle (2qp) energy as
$E_{\alpha}+E_{\beta} \leq 60$ MeV.
This reduces the number of 2qp states to, for instance,
about 38 000 for the $K^{\pi}=0^{-}$ excitation in $^{154}$Sm.
The calculation of the QRPA matrix elements in the qp basis 
is performed in the parallel computers. 
In the present calculation, all the matrix elements are real and 
we use 512 processors to compute them.  
The two-dimensional block cyclic distribution is employed to 
keep a good load balancing.

To save the computing time for diagonalization of the QRPA matrix, 
we employ a technique to reduce the non-Hermitian eigenvalue problem to 
a real symmetric matrix of half the dimension~\cite{ull71,RS}. 
For diagonalization of the matrix, 
we use the ScaLAPACK {\sc pdsyev} subroutine~\cite{scalapack}.
To calculate the QRPA matrix elements and to diagonalize the matrix, 
it takes about 390 CPU hours and 135 CPU hours, respectively 
on the RICC, the supercomputer facility at RIKEN.  

The similar calculations of the HFB+QRPA for axially deformed nuclei  
have been recently reported~\cite{per08,pen09,los10,ter10}.
Among them, the one by Terasaki and Engel in Ref.~\cite{ter10} 
is analogous to ours. 
They adopt the canonical-basis representation and introduce a further truncation 
according to the occupation probabilities of 2qp excitations. 
In contrast, we adopt the qp representation and truncation simply due to the 2qp energies. 
However, we have a drawback in the computing time. 
Carrying out the numerical integration for the p-h matrix elements 
in the qp basis takes 4 times as long as 
the calculation in the canonical basis.
For reference, we show the matrix elements of the QRPA 
in the qp basis in Appendix A. 

Since the full self-consistency between the static mean-field
calculation and the dynamical calculation is slightly violated by
neglecting two-body Coulomb interaction and truncating the 2qp space,
the spurious states may have finite excitation energies. 
In the present calculation, the spurious states 
for the $K^{\pi}=0^{+}, 1^{+}, 0^{-}$ and $1^{-}$ excitations 
appear at 0.35 MeV, 0.34 MeV, 1.46$i$ MeV and 1.60$i$ MeV, 
respectively in $^{154}$Sm.
We see in section III.B 
the contamination of the spurious component in GRs to be small 
because the GRs are well apart from the spurious states in energy.

The transition strength distribution as a function of the 
excitation energy $E$ is calculated as
\begin{equation}
\label{S_l}
S_{\lambda}^{\tau}(E)
=\sum_{i}\sum_{K} \dfrac{\gamma/2}{\pi}\dfrac{|\langle i|\hat{F}_{\lambda K}^{\tau}|0\rangle|^{2}}
{(E-\hbar \omega_{i})^{2}+\gamma^{2}/4}.
\end{equation} 
The smearing width $\gamma$ is set to 2 MeV, which is supposed to simulate
the spreading effect, $\Gamma^\downarrow$, missing in the QRPA. 
It is noted that in Ref.~\cite{yos11} we showed that the constant smearing parameter 
of $\gamma=2$ MeV reproduces well the total width of the GDR in the Nd and Sm isotopes 
with $N=82 - 92$. 

Here we define the operators as
\allowdisplaybreaks[1]
\begin{align}
\hat{F}^{\tau=0}_{\lambda=0} &=
\sum_{\tau_{3}=1,-1} \int d\boldsymbol{r} r^{2}\hat{\psi}_{\tau_{3}}^{\dagger}(\boldsymbol{r})
\hat{\psi}_{\tau_{3}}(\boldsymbol{r}), \\
\hat{F}^{\tau=1}_{\lambda=0} &=
\sum_{\tau_{3}=1,-1} \int d\boldsymbol{r} \tau_{3}r^{2}\hat{\psi}_{\tau_{3}}^{\dagger}(\boldsymbol{r})
\hat{\psi}_{\tau_{3}}(\boldsymbol{r}), \\
\hat{F}^{\tau=0}_{\lambda=1,K} &=
\sum_{\tau_{3}=1,-1} \int d\boldsymbol{r} r^{3}Y_{1 K}(\hat{r})
\hat{\psi}_{\tau_{3}}^{\dagger}(\boldsymbol{r})
\hat{\psi}_{\tau_{3}}(\boldsymbol{r}), \label{op_IS_dip}\\
\hat{F}^{\tau=1}_{\lambda=1,K} &=
\int  d\boldsymbol{r}rY_{1K}(\hat{r})\left\{
\dfrac{Z}{A}\hat{\psi}_{\nu}^{\dagger}(\boldsymbol{r})\hat{\psi}_{\nu}(\boldsymbol{r})
-\dfrac{N}{A}\hat{\psi}_{\pi}^{\dagger}(\boldsymbol{r})\hat{\psi}_{\pi}(\boldsymbol{r})
\right \}
, \\ 
\hat{F}^{\tau=0}_{\lambda=2,K} &= \sum_{\tau_{3}=1,-1} \int d\boldsymbol{r} r^{2}Y_{2 K}(\hat{r})
\hat{\psi}_{\tau_{3}}^{\dagger}(\boldsymbol{r})
\hat{\psi}_{\tau_{3}}(\boldsymbol{r}), \\
\hat{F}^{\tau=1}_{\lambda=2,K} &=
\sum_{\tau_{3}=1,-1} \int d\boldsymbol{r} \tau_{3}r^{2}Y_{2 K}(\hat{r})
\hat{\psi}_{\tau_{3}}^{\dagger}(\boldsymbol{r})
\hat{\psi}_{\tau_{3}}(\boldsymbol{r}), \\
\hat{F}^{\tau=0}_{\lambda=3,K} &=
\sum_{\tau_{3}=1,-1} \int d\boldsymbol{r} r^{3}Y_{3 K}(\hat{r})
\hat{\psi}_{\tau_{3}}^{\dagger}(\boldsymbol{r})
\hat{\psi}_{\tau_{3}}(\boldsymbol{r}), \\
\hat{F}^{\tau=1}_{\lambda=3,K} &=
\sum_{\tau_{3}=1,-1} \int d\boldsymbol{r} \tau_{3}r^{3}Y_{3 K}(\hat{r})
\hat{\psi}_{\tau_{3}}^{\dagger}(\boldsymbol{r})
\hat{\psi}_{\tau_{3}}(\boldsymbol{r}).
\end{align} 
The spin index is omitted for simplicity in the above definition 
because the spin direction is unchanged by these operators. 

\section{Results and discussion}\label{result}

\begin{table*}[t]
\begin{center}
\caption{Ground state properties of Nd and Sm isotopes 
obtained by the deformed HFB calculation 
with the SkM* and pairing EDFs. 
Chemical potentials $\lambda_q$, deformation parameters $\beta_2^q$,
quadrupole moments $\langle Q_{2} \rangle_q$, 
average pairing gaps $\langle \Delta \rangle_q$,
and root-mean-square radii $\sqrt{\langle r^{2} \rangle_q}$ for 
neutrons and protons are listed.}
\label{GS}
\begin{tabular}{crrrrrrrrrrrr}
\hline \hline
\noalign{\smallskip}
 & $^{142}$Nd  & $^{144}$Nd & $^{146}$Nd & $^{148}$Nd 
 & $^{150}$Nd & $^{152}$Nd 
 & $^{144}$Sm  & $^{146}$Sm & $^{148}$Sm & $^{150}$Sm
 & $^{152}$Sm & $^{154}$Sm \\
\noalign{\smallskip}\hline\noalign{\smallskip}
$\lambda_{\nu}$ (MeV)  & $-8.79$  & $-6.42$ & $-6.65$  &  $-6.73$
& $-7.03$ & $-6.69$ 
& $-9.12$  & $-6.99$ & $-7.21$  & $-7.42$
& $-7.58$ & $-7.25$ \\
$\lambda_{\pi}$ (MeV)  & $-5.23$ & $-5.83$ & $-6.25$ & $-6.90$
& $-7.72$ & $-8.46$ 
& $-4.39$  & $-4.99$ & $-5.37$ & $-5.99$
& $-6.60$ & $-7.13$\\
$\beta_{2}^{\nu}$  & 0.00  & 0.00 & 0.12 & 0.18
& 0.26 & 0.30 
& 0.00 & 0.00  & 0.12 & 0.20
& 0.27 & 0.30 \\
$\beta_{2}^{\pi}$  & 0.00 & 0.00 & 0.14 & 0.21
& 0.30 & 0.34 
& 0.00 & 0.00 & 0.14 & 0.22
& 0.30 & 0.33 \\
$\langle Q_{2} \rangle_{\nu}$ (fm$^{2}$)  & $\sim 0$ & $\sim 0$ & 328 & 530
& 796 & 939  
& $\sim 0$ & $\sim 0$ & 323 & 563
& 805 & 939 \\
$\langle Q_{2} \rangle_{\pi}$ (fm$^{2}$) & $\sim 0$ & $\sim 0$ & 251 & 389
& 568 & 644 
& $\sim 0$ & $\sim 0$ & 257  & 435
& 597 & 668 \\
$\langle \Delta \rangle_{\nu}$ (MeV) & 0.00  & 0.82 & 0.93 & 1.06
& 0.99 & 0.78 
& 0.00 & 0.86 & 0.98 & 1.10
& 1.07 & 0.90 \\
$\langle \Delta \rangle_{\pi}$ (MeV) & 1.71 & 1.67 & 1.48 & 1.30
& 0.87 & 0.54
& 1.75 & 1.72 & 1.57 & 1.35
& 1.04 & 0.90  \\
$\sqrt{\langle r^{2} \rangle_{\nu}}$ (fm)  & 4.95  & 4.99 & 5.03 & 5.08
& 5.15 & 5.20
& 4.97 & 5.00 & 5.04 & 5.10
& 5.16 & 5.20 \\
$\sqrt{\langle r^{2} \rangle_{\pi}}$ (fm)  & 4.86  & 4.87 & 4.90 & 4.93
& 4.99 & 5.02 
& 4.89 & 4.90 & 4.93 & 4.98
& 5.02 & 5.06 \\
\noalign{\smallskip}
\hline \hline
\end{tabular}
\end{center}
\end{table*}

\subsection{Ground state properties}\label{GS_property}
We summarize in Table~\ref{GS} the calculated 
ground-state properties of the Nd and Sm isotopes.
Around $N=82$, the systems are calculated to be spherical. 
The calculated quadrupole moment 
of $^{142, 144}$Nd and $^{144, 146}$Sm are very small but finite. 
This is due to
the numerical error originating from the finite mesh size 
and breaking of the spherical symmetry of the rectangular box employed. 
Increase in the neutron number, the deformation gradually develops.
As shown in Fig.~1 of Ref.~\cite{yos11}, the calculation well reproduces 
the evolution of quadrupole deformation for $N \ge 86$. 

The pairing gap disappears at $N=82$ associated with the spherical magic number 
of neutrons. 
The obtained pairing gaps are in good agreement with the empirical values
for deformed nuclei,
while they are overestimated in the spherical systems.
This is consistent with the findings of Ref.~\cite{ber09} 
that the pairing gaps of deformed nuclei 
are underestimated if we use the pairing functional adjusted to
the experimental 
data for spherical nuclei. 
Note that the pairing functional employed in the present calculation is constructed 
by adjusting to the experimental pairing gaps of deformed nuclei~\cite{yam09}. 

\subsection{Mixing of spurious center-of-mass motion}

\begin{figure}[t]
\begin{center}
\includegraphics[scale=0.65]{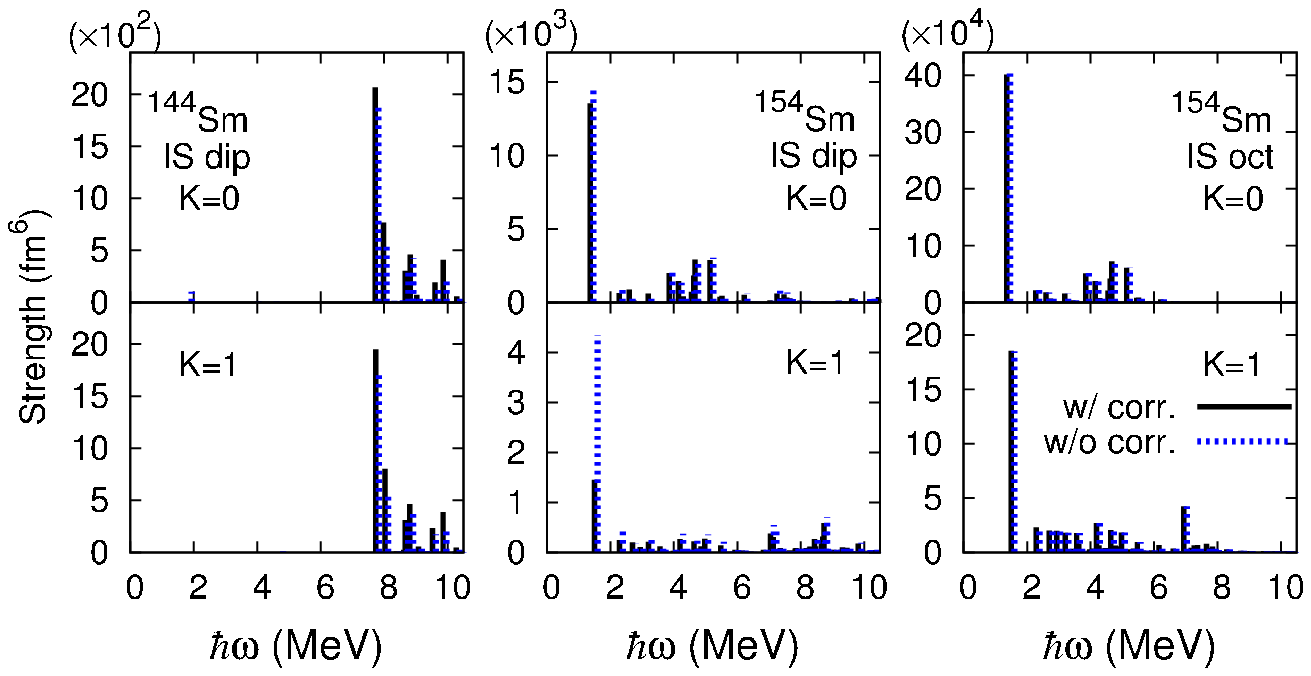}
\includegraphics[scale=0.65]{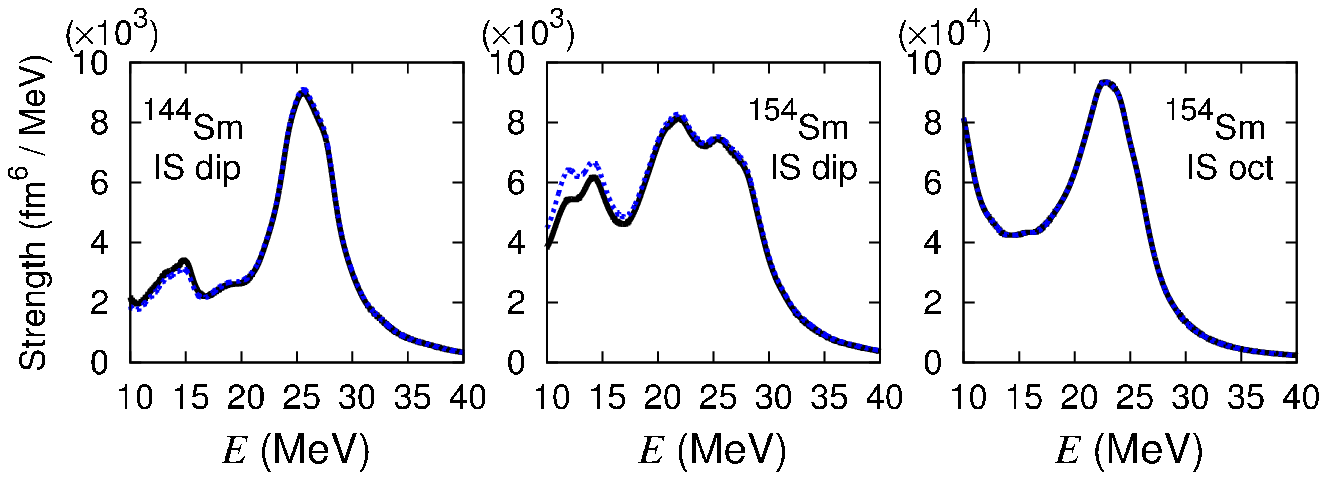}
\caption{(Color online) 
The IS dipole and octupole transition-strength distributions in  $^{144}$Sm and 
in $^{154}$Sm 
in the low-energy region (upper panels) and the GR energy region (lower).
The results obtained with $\eta_K=0$ and $\eta_K^{(3)}=0$ are shown by
dotted lines.
}
\label{IS_spurious}
\end{center}
\end{figure}

The isoscalar (IS) dipole operator, Eq.~(\ref{op_IS_dip}) contains the 
component of the center-of-mass motion.
For deformed nuclei,
the $K^{\pi} = 0^-$ and $1^-$ octupole operators may
also excite the spurious center-of-mass motion. 
To examine the mixing of the spurious modes,
we use the corrected operator;
\begin{equation}
\hat{F}^{\tau=0}_{\lambda=1,K} =
\sum_{\tau_{3}} \int d\boldsymbol{r} (r^{3}-\eta_{K}r)Y_{1 K}(\hat{r})
\hat{\psi}_{\tau_{3}}^{\dagger}(\boldsymbol{r})
\hat{\psi}_{\tau_{3}}(\boldsymbol{r}) \label{op_IS_dip_cor}
\end{equation}
instead of using Eq.~(\ref{op_IS_dip}). 
Here, the correction factor in the isoscalar dipole operator 
originally discussed for a spherical
system ($\eta$) to subtract the spurious component of the center-of-mass 
motion~\cite{gia81} was extended to a deformed system ($\eta_{K}$)~\cite{yos08}, and
coincides with $\eta_{K}=\eta=5/3$ in the spherical limit.
For the octupole operators,
we use a similar technique to the case of 
the dipole operator~\cite{yos09b};
\begin{equation}
\hat{F}^{\tau=0}_{\lambda=3,K} =
\sum_{\tau_{3}} \int d\boldsymbol{r} [r^{3}Y_{3 K}(\hat{r})-\eta^{(3)}_{K}rY_{1 K}(\hat{r})]
\hat{\psi}_{\tau_{3}}^{\dagger}(\boldsymbol{r})
\hat{\psi}_{\tau_{3}}(\boldsymbol{r}). \label{op_IS_oct_cor}
\end{equation}
It is noted that the correction factor $\eta^{(3)}_K$ vanishes in the spherical limit.

In Fig.~\ref{IS_spurious} we show
the IS dipole and octupole 
transition-strength distributions in the low-energy region 
in $^{144}$Sm and $^{154}$Sm,
calculated with and without the correction terms, $\eta_K$ and $\eta_K^{(3)}$. 
In $^{144}$Sm, because the transition strengths calculated with finite
$\eta_K$ are approximately identical to those with $\eta_K=0$,
the low-energy dipole states around 8 MeV
are almost free from the spurious center-of-mass motion.
However, for the lowest $K=1$ dipole state in $^{154}$Sm,
we see a large difference between the two calculations.
This implies that the full self-consistency is necessary to 
describe quantitatively the low-lying dipole states.
The contamination of the spurious mode is smaller 
in the low-lying octupole excitations and in the GRs 
as shown in the lower panel of Fig.~\ref{IS_spurious}. 

\subsection{Giant resonances}

Let us discuss properties of GRs.
In order to quantify the excitation energy of the GR,
two kinds of definition are utilized.
The centroid energy $E_c$ is frequently used in the experimental analysis,
defined by
\begin{equation}
E_c=\dfrac{m_1}{m_0},
\end{equation}
where $m_k$ is a $k-$th moment of 
the transition strength distribution in an energy interval of $[E_a, E_b]$ MeV.
\begin{equation}
\label{m_k_con}
m_k  \equiv  \int_{E_a}^{E_b} E^k S_\lambda^\tau (E)  dE, 
\end{equation}
where $S^\tau_\lambda (E)$ is defined by 
Eq.~(\ref{S_l}) in the calculation. 
We take the upper and lower limits, $[E_a, E_b]$,
same as those used in the experimental analysis.

Another definition of the excitation energy is denoted as $E_x$.
This is extracted by fitting the strength distribution of the GR,
$S_\lambda^\tau (E)$,
by the Lorentz curve with two parameters, the peak energy $E_x$ and
the width $\Gamma$.

\subsubsection{Positive-parity excitation}\label{positive}

\begin{figure}[t]
\begin{center}
\includegraphics[scale=0.33]{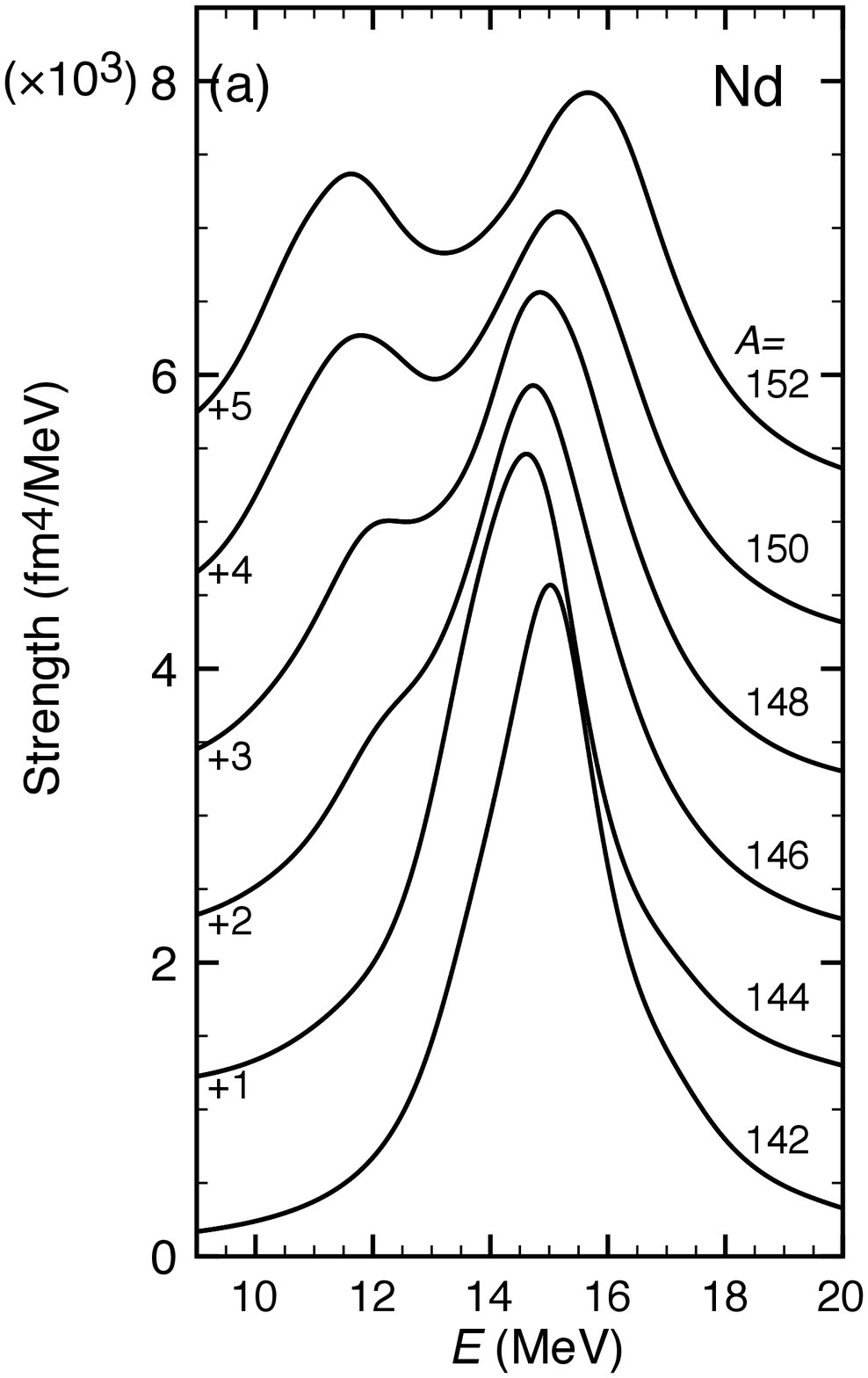}
\includegraphics[scale=0.33]{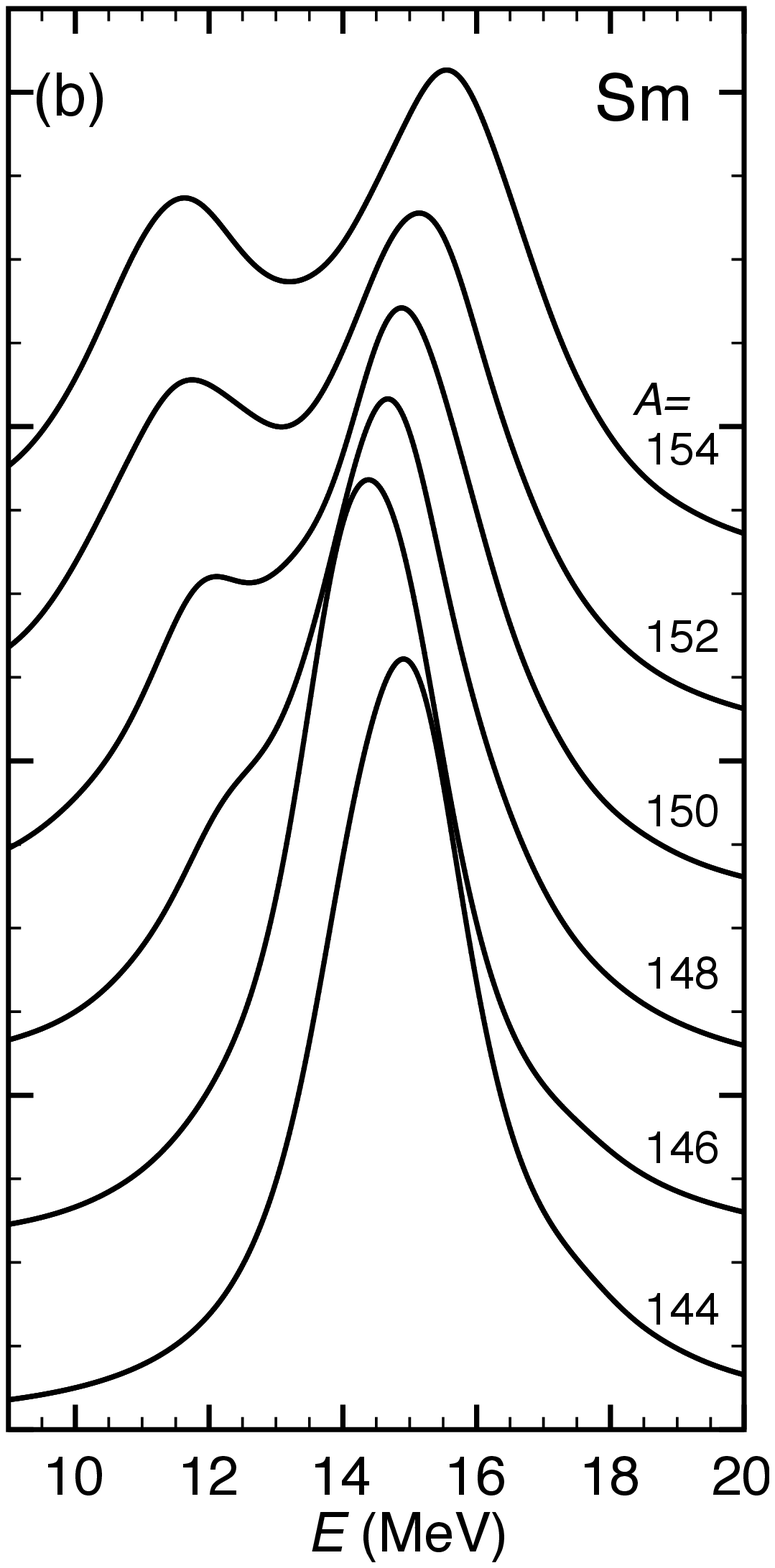} \\
\includegraphics[scale=0.33]{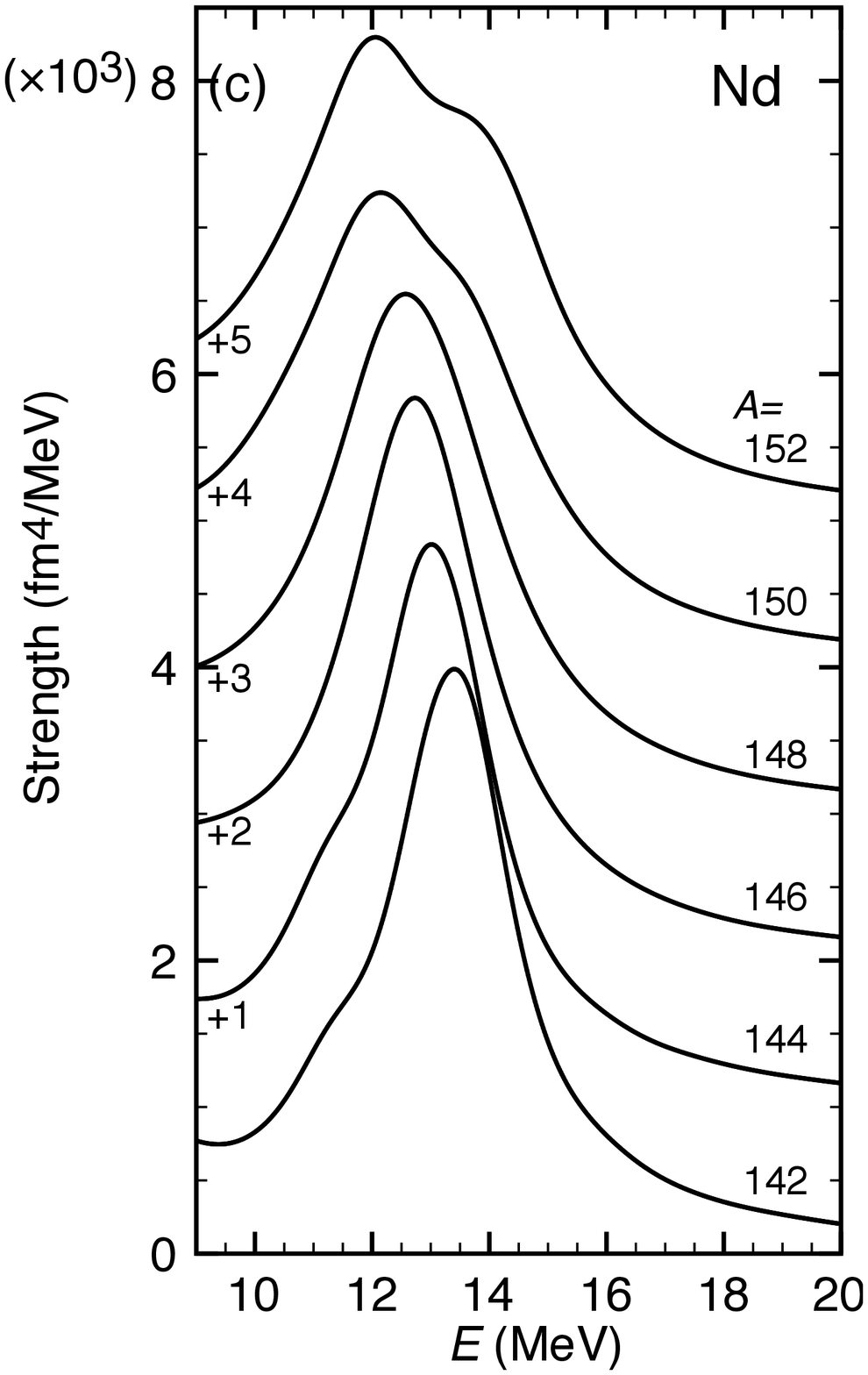}
\includegraphics[scale=0.33]{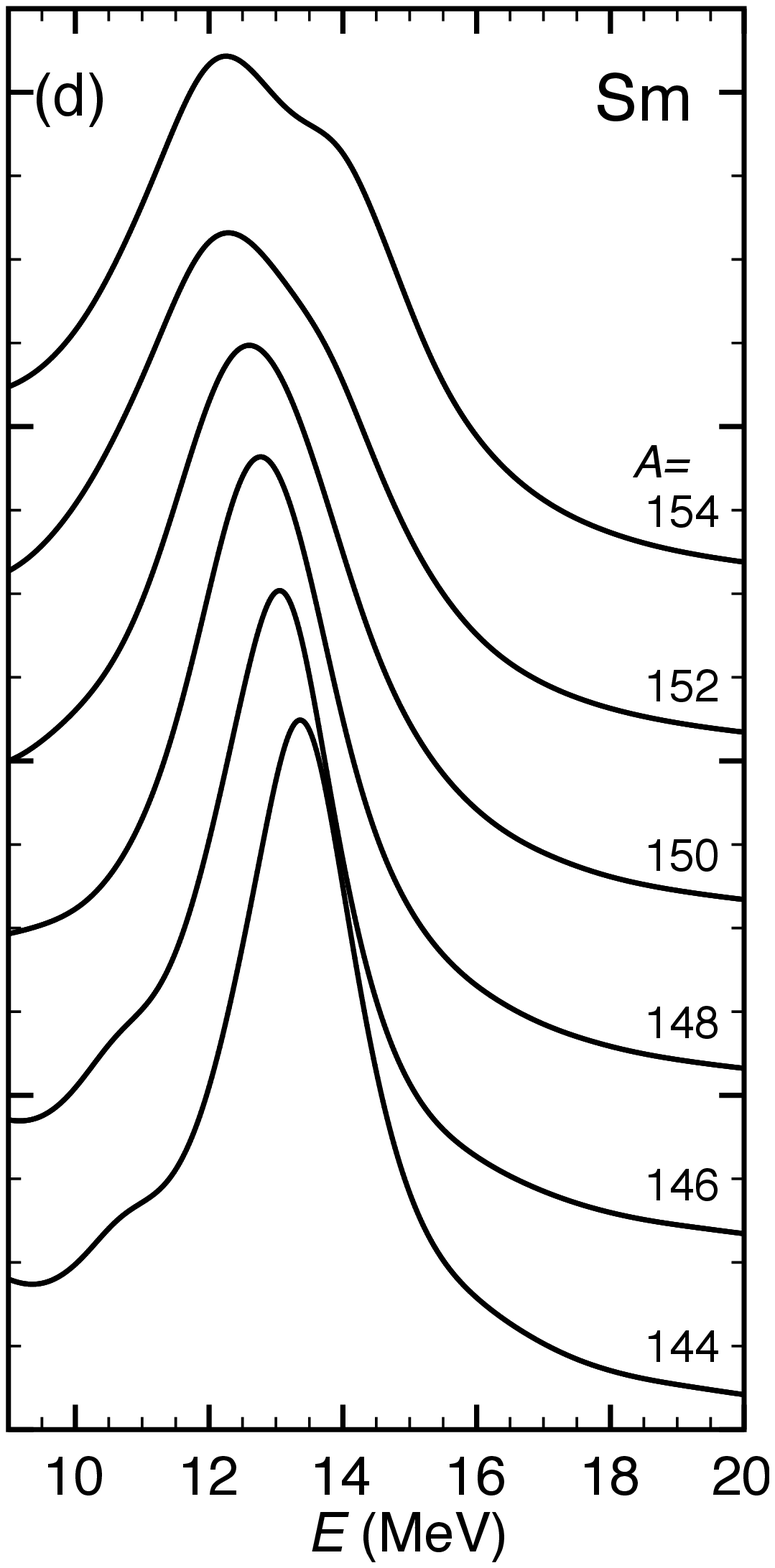}
\caption{The strength distributions (shifted) of ISGMR [(a), (b)] and ISGQR [(c), (d)] 
in Nd and Sm isotopes.}
\label{IS_positive}
\end{center}
\end{figure}

\begin{figure}[t]
\begin{center}
\includegraphics[scale=0.27]{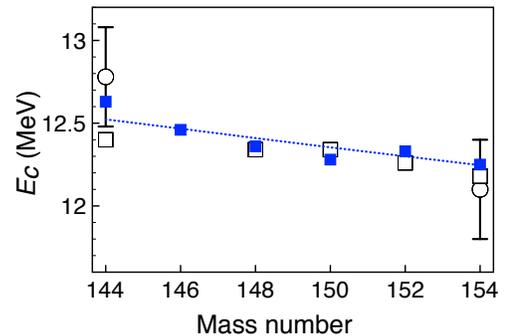}
\caption{(Color online) 
The centroid energies of the ISGQR in the Sm isotopes with a fitted line. 
The experimental data~\cite{itoh, you04} 
are denoted by open symbols.}
\label{Sm_ISGQR_energy}
\end{center}
\end{figure}

Figure~\ref{IS_positive} shows the strength distributions of 
IS monopole and quadrupole excitations in the Nd and Sm 
isotopes. 
We discuss first the giant quadrupole resonance (GQR). 
Both in the Nd and Sm isotopes, 
ISGQRs are located around 12--14 MeV. 
With increase in the mass number, 
the peak energy of the ISGQR becomes smaller. 
This is consistent with the experiment on 
the systematic observation of the ISGQR energy in the Sm isotopes~\cite{ito03, you04}.
Figure \ref{Sm_ISGQR_energy} shows the 
centroid energy of the ISGQR in the Sm isotopes.
Here we used 
the energy interval of [9,15] MeV.
Open squares in Fig.~\ref{Sm_ISGQR_energy} are
obtained from the strength distribution in Ref.~\cite{ito03}. 
The present results well reproduce the experimental data.
The calculated centroid energy is well fitted by 
the $65.6 A^{-1/3}$ line,
which agrees with the empirical behavior,
$(64.0 \pm 1.7) \times A^{-1/3}$~\cite{har01}.
Dependence on the choice of the Skyrme functional is discussed later.

The ISGQR in spherical nuclei was successfully described by
the pairing-plus-quadrupole (P+Q) model.
However, the same model failed to reproduce the observed data in
deformed nuclei.
This failure can be attributed to the violation of the
nuclear self-consistency between the shapes of the potential 
and the density distributions.
Making use of the quadrupole operator in doubly-stretched coordinates
significantly improves the results~\cite{kis75}.
It is due to the fact that the doubly-stretched quadrupole operator 
appropriately describes the quadrupole fluctuation about a
deformed ground state~\cite{sak89a}.
In fact, the predicted deformation splitting of the ISGQR in $^{154}$Sm 
is calculated to be about 2 MeV using the doubly-stretched
P+Q model, whereas it is about 6 MeV
using the ordinary P+Q model.

\begin{figure}[t]
\begin{center}
\includegraphics[scale=0.27]{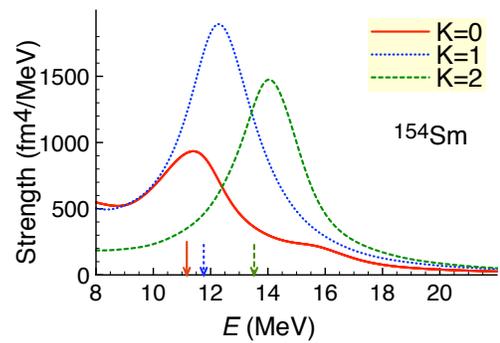}
\caption{(Color online) The IS quadrupole transition-strength distribution 
in $^{154}$Sm for the $K^{\pi}=0^{+}$, 
$1^{+}$, and $2^{+}$ excitations. 
The eigenenergies obtained with use of the doubly-stretched P+Q model 
are denoted by the arrows, 
and the peak position of the GQR was adjusted to the 
experimental data~\cite{kis75}. }
\label{154Sm_IS2+}
\end{center}
\end{figure}

Figure~\ref{154Sm_IS2+} shows the IS quadrupole transition-strength distribution 
in $^{154}$Sm for the $K^{\pi}=0^{+}$, $1^{+}$ and $2^{+}$ excitations.
The $K$ splitting, $E_{K=2}-E_{K=0}$,
for the ISGQR is 2.8 MeV in the present calculation. 
This is consistent with the value obtained in the doubly-stretched P+Q model 
and the experimental observation~\cite{kis75}. 
This indicates that the present calculation based 
on the EDF naturally takes into account the nuclear self-consistency, 
which has to be introduced explicitly in the P+Q model 
where the higher-order terms are required additionally to satisfy the 
nuclear self-consistency~\cite{sak89}. 
Since the energy splitting associated with the deformation 
is comparable to the smearing parameter, 
the deformation splitting, which is clearly visible in the photoabsorption 
cross sections~\cite{yos11}, does not appear in the ISGQR.
Instead, we find a broadening of the width for the ISGQR 
associated with the development of the deformation 
(see the table in Appendix B). 

Next, let us discuss the monopole excitation. 
In the spherical nuclei, we can see a sharp peak at around 15 MeV 
which is identified as the ISGMR. 
In $^{144}$Sm, the peak energy and the width are $E_{x}=14.8$ MeV and $\Gamma=2.61$ MeV. 
This is compatible with the observed values of $E_{x}=15.40 \pm 0.30$ and 
$\Gamma=3.40 \pm 0.20$ MeV~\cite{you04}. 

The ISGMR in deformed nuclei has a double-peak structure. 
The lower-energy peak ($8 < E < 13.5$ MeV) 
and the higher-energy peak ($13.5 < E < 19$ MeV) 
exhaust $31.4 \%$ and $60.6 \%$ of the IS monopole energy-weighted-sum rule (EWSR) 
value, $3.38 \times 10^{5}$ fm$^{4}$ MeV, in $^{154}$Sm. 
The higher-energy peak of the IS monopole strength is identified 
as a primal ISGMR and the lower-energy peak is associated with the coupling 
to the $K^{\pi}=0^{+}$ component of the ISGQR. 
The lower peak of the ISGMR around 11 MeV,
is located at the peak position of the $K^{\pi}=0^{+}$ component of 
the ISGQR  shown in Fig.~\ref{154Sm_IS2+}.

\begin{figure}[t]
\begin{center}
\includegraphics[scale=0.27]{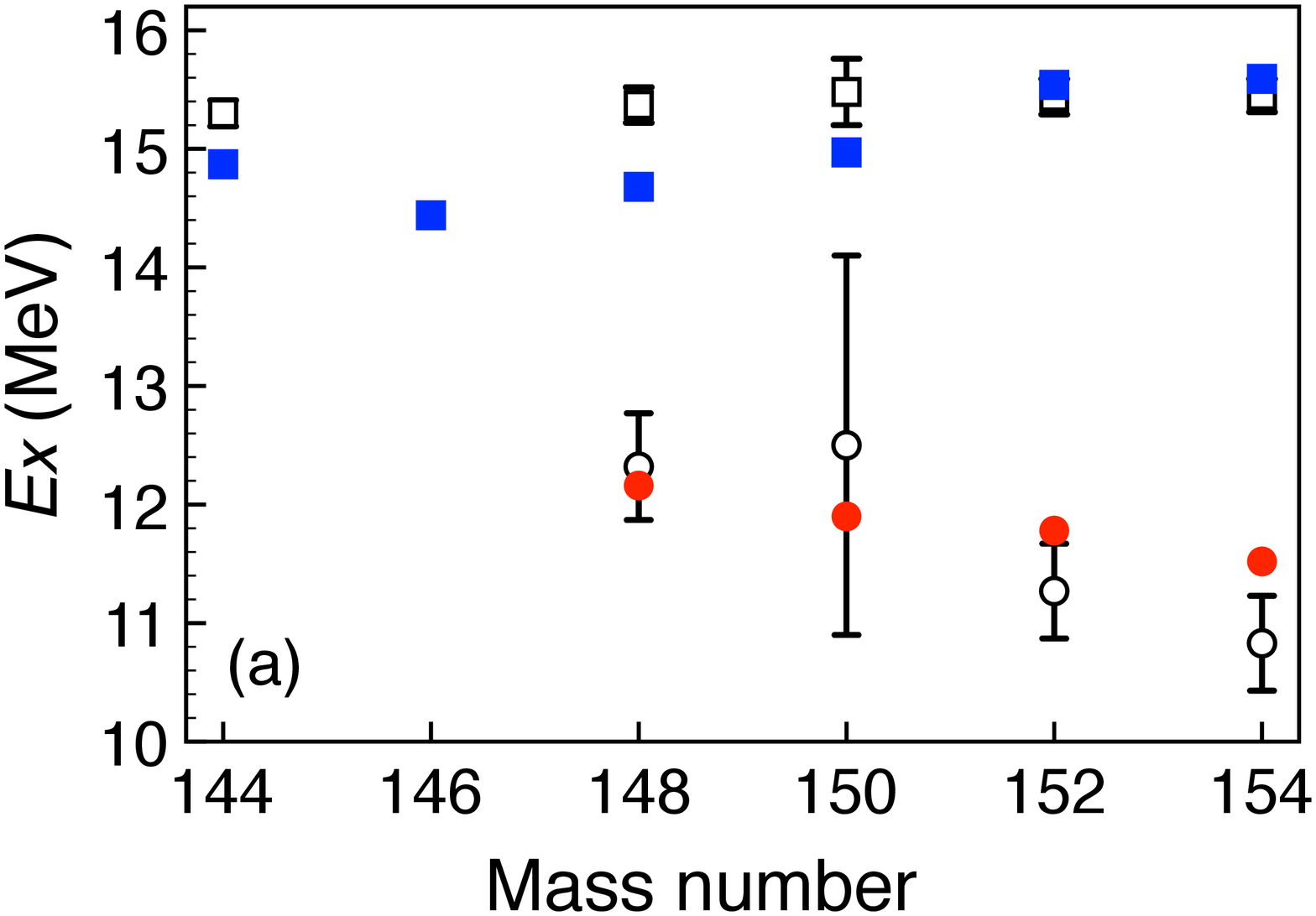}
\includegraphics[scale=0.27]{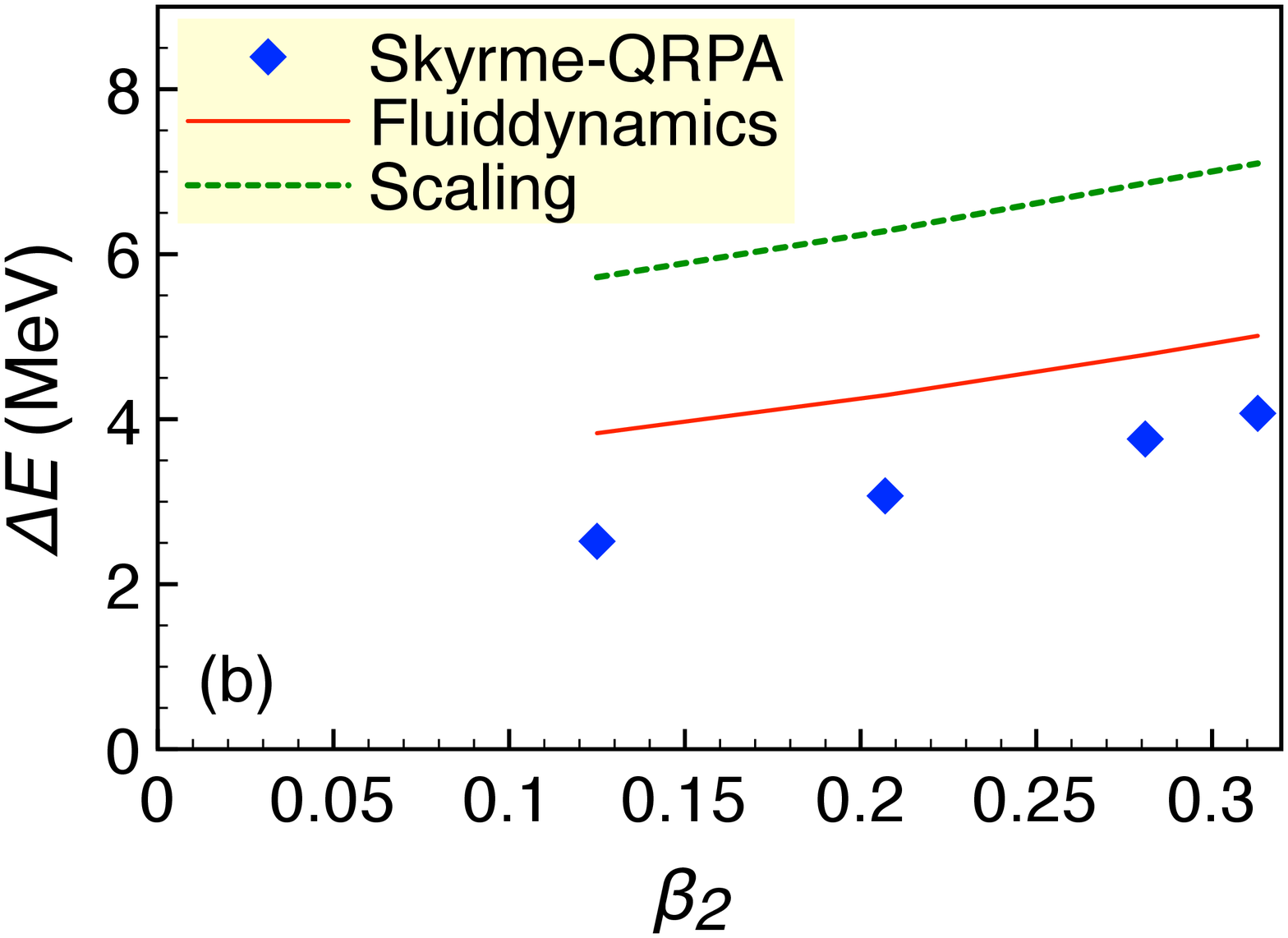}
\caption{(Color online) (a)The excitation energies of the ISGMR in the Sm isotopes.  
The experimental data~\cite{ito03} are denoted by open symbols with error bars. 
(b) The energy difference of the upper peak and the lower peak of the ISGMR 
in $^{148,150,152,154}$Sm as a function of the deformation parameter $\beta_2$. 
The lines are results of the fluiddynamical and scaling models~\cite{nis85}. }
\label{Sm_ISGMR_energy}
\end{center}
\end{figure}

\begin{table*}[t]
\begin{center}
\caption{The parameters of the ISGMR in $^{154}$Sm. 
Fitting the strength function $S_0^0(E)$ with $\gamma=2$ MeV 
by a sum of two Lorentz lines,
the peak energy $E_{x}$ and the width $\Gamma$ are extracted.
The energy-weighted sum (EWS) of the transition strength is calculated 
in the energy range of [8,13.5] MeV and [13.5,19] MeV for the lower peak and upper peak 
of the ISGMR in the calculation using the SkM* functional. 
The energy range is slightly changed according to 
the shift of peak positions of the ISGMR 
in the calculations using the SLy4 and SkP functionals. 
The experimental values are taken from Refs.~\cite{you04} and \cite{ito02}. 
Results of other calculations employing different kinds of model 
in Refs.~\cite{nis85} and \cite{abg80} and are also included.}
\label{154Sm_GMR}
\begin{tabular}{cccccccc}
\hline \hline
\noalign{\smallskip}
 & & Lower peak  & & & Upper peak &  & Ratio of EWS\\
 & $E_{x}$ & $\Gamma$ & EWS & $E_{x}$ & $\Gamma$ & EWS & Upper/Lower  \\
 & (MeV)  & (MeV) & $\%$ & (MeV) & (MeV) & $\%$  & \\
\noalign{\smallskip}\hline\noalign{\smallskip}
SkM*  & 11.5 & 3.75 & 31.4 & 15.6 & 2.73 & 60.6 & 1.9  \\
SLy4  & 12.1 & 3.62 & 36.3 & 16.2 & 2.68 & 57.0 & 1.6 \\
SkP  & 10.3 & 3.48 & 21.8 & 14.7 & 2.78 & 70.8 & 3.2  \\
TAMU~\cite{you04} & $11.05 \pm 0.05$ & $3.2\pm0.1$ & $32\pm2$ & $15.17\pm0.05$ & $4.0\pm0.1$ & $80\pm5$ & $2.5 \pm 0.2$ \\
RCNP~\cite{ito02} & $11.0 \pm 0.8$ & (5.1) & $17.5 \pm 5$ & $15.6 \pm 0.2$ & (3.9) & $69\pm5$ & $3.9 \pm 1.2$ \\
Fluiddynamics~\cite{nis85} & 10.1 &  & 21.5 & 15.6 &  & 76.3 & 3.5 \\
Scaling~\cite{nis85} & 11.0 &  &  16.6 & 18.1 & & 83.4 & 5.0 \\  
Cranking~\cite{abg80} & 10.4 &  & 21 & 15.9 & & 79 & 3.8 \\
\noalign{\smallskip}
\hline \hline
\end{tabular}
\end{center}
\end{table*}

Figure \ref{Sm_ISGMR_energy}(a) shows the peak energy of the ISGMR in 
the Sm isotopes.  
The calculation shows an excellent agreement with
the experimental data both in spherical and deformed nuclei.
As the deformation develops from $^{148}$Sm, 
the higher-energy peak of the ISGMR slightly increases. 
In Fig.~\ref{Sm_ISGMR_energy}(b), 
the energy difference of the upper and lower peaks
of the ISGMR is shown as a function of the 
deformation parameter of the ground state. 
The results are compared with the predictions by
the fluiddynamics model and the simple scaling model with
the effective mass $m^*/m=0.8$ and the Landau parameter $F_0=-0.25$~\cite{nis85}.
The result of the fluiddynamics model is consistent with our result,
although it underestimates the excitation energy of the low-energy
peak of ISGMR.
The deformation dependence of the splitting energy is well reproduced.
On the other hand, the simple scaling model significantly
overestimates the ISGMR peak energy,
which results in too large splitting of the peak energies.

Since the experimental studies for the detailed structure of the ISGMR in $^{154}$Sm 
are available~\cite{you04,ito02}, we are going to discuss here 
the properties of the calculated ISGMR in $^{154}$Sm. 
Table~\ref{154Sm_GMR} summarizes the parameters of the ISGMR in $^{154}$Sm. 
The peak energy $E_{x}$ and the width $\Gamma$ in a deformed system 
are obtained by fitting the strength distribution with a sum of two Lorentz lines. 
The calculations are compared with inelastic $\alpha$ scattering experiments 
at Texas A$\&$M University~\cite{you04} and at RCNP, Osaka University~\cite{ito02}. 
Results of the calculations employing the SLy4~\cite{cha98} and SkP~\cite{dob84} 
functionals and other models~\cite{abg80,nis85} are also shown. 
The same pairing energy functional, Eq. (\ref{pair_int}),
is used in all the calculations.

The excitation energies are described best by the 
SkM* functional among three kinds of functionals. 
The ratio of the energy-weighted sum of the strengths for the upper peak
to that for the lower peak 
varies from 1.6 (SLy4) to 3.2 (SkP), 
and the SkP gives better agreement with the experimental data.
This implies that the coupling effect between the GMR and the GQR is weaker 
for the SkP functional than for the SkM* and SLy4 functionals. 
As discussed above, the coupling is determined by the quadrupole moment 
(deformation parameter) of the ground state. 
Indeed, the mass deformation parameter obtained in the present calculation 
is $\beta_{2}=0.29$ for SkP, while $\beta_{2}=0.31$ for SkM* and SLy4. 

\begin{figure}[t]
\begin{center}
\includegraphics[scale=0.33]{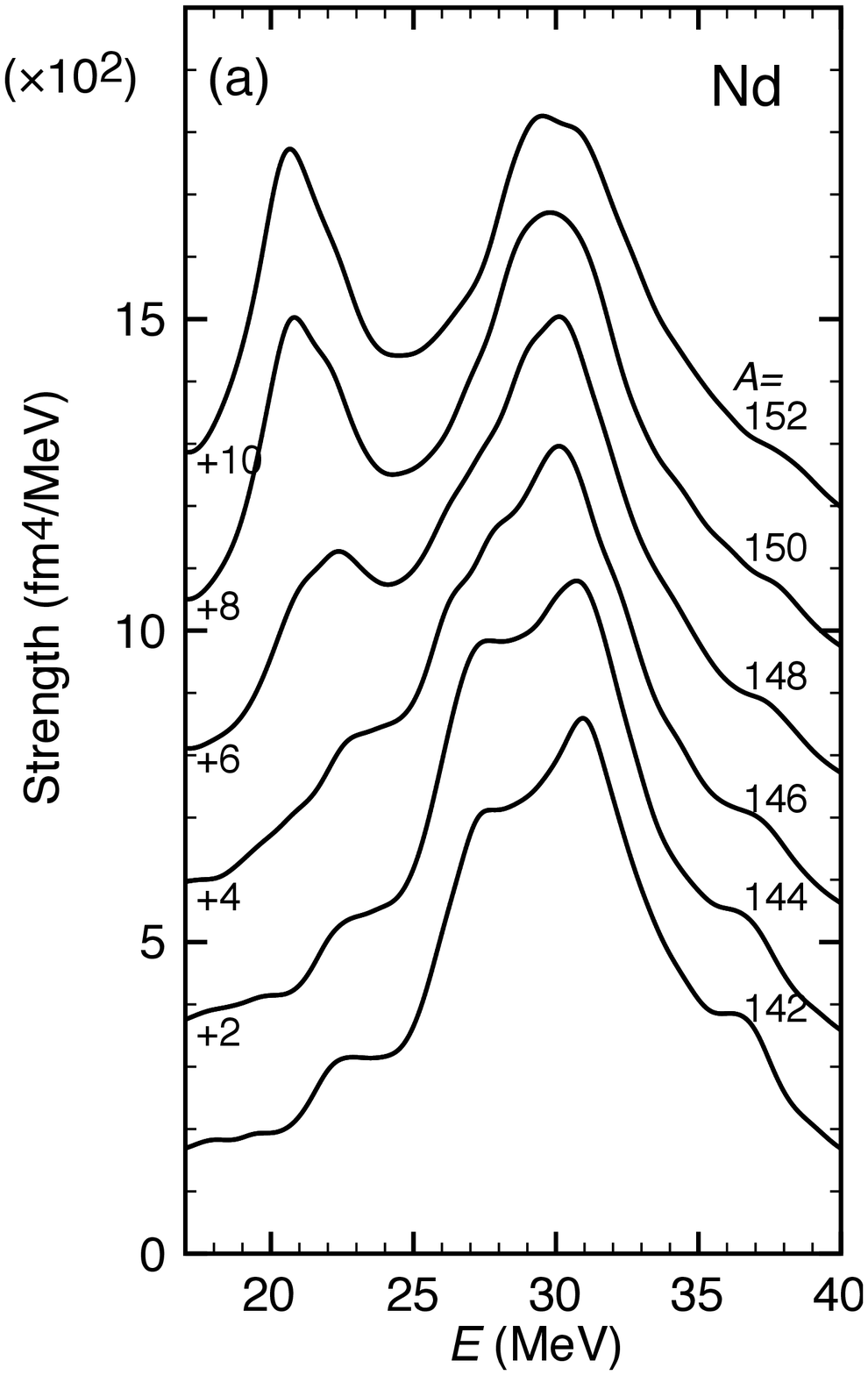}
\includegraphics[scale=0.33]{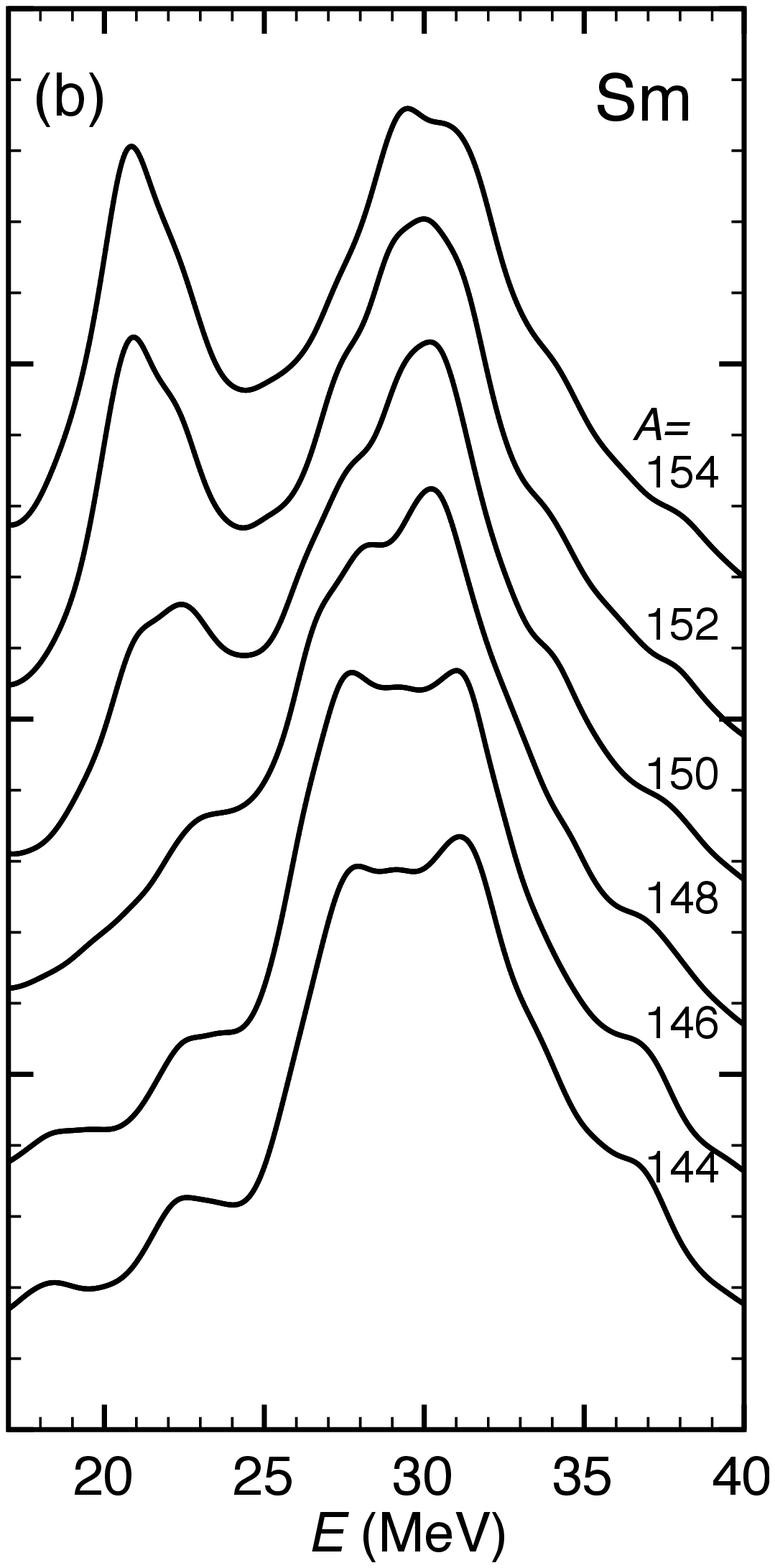} \\
\includegraphics[scale=0.33]{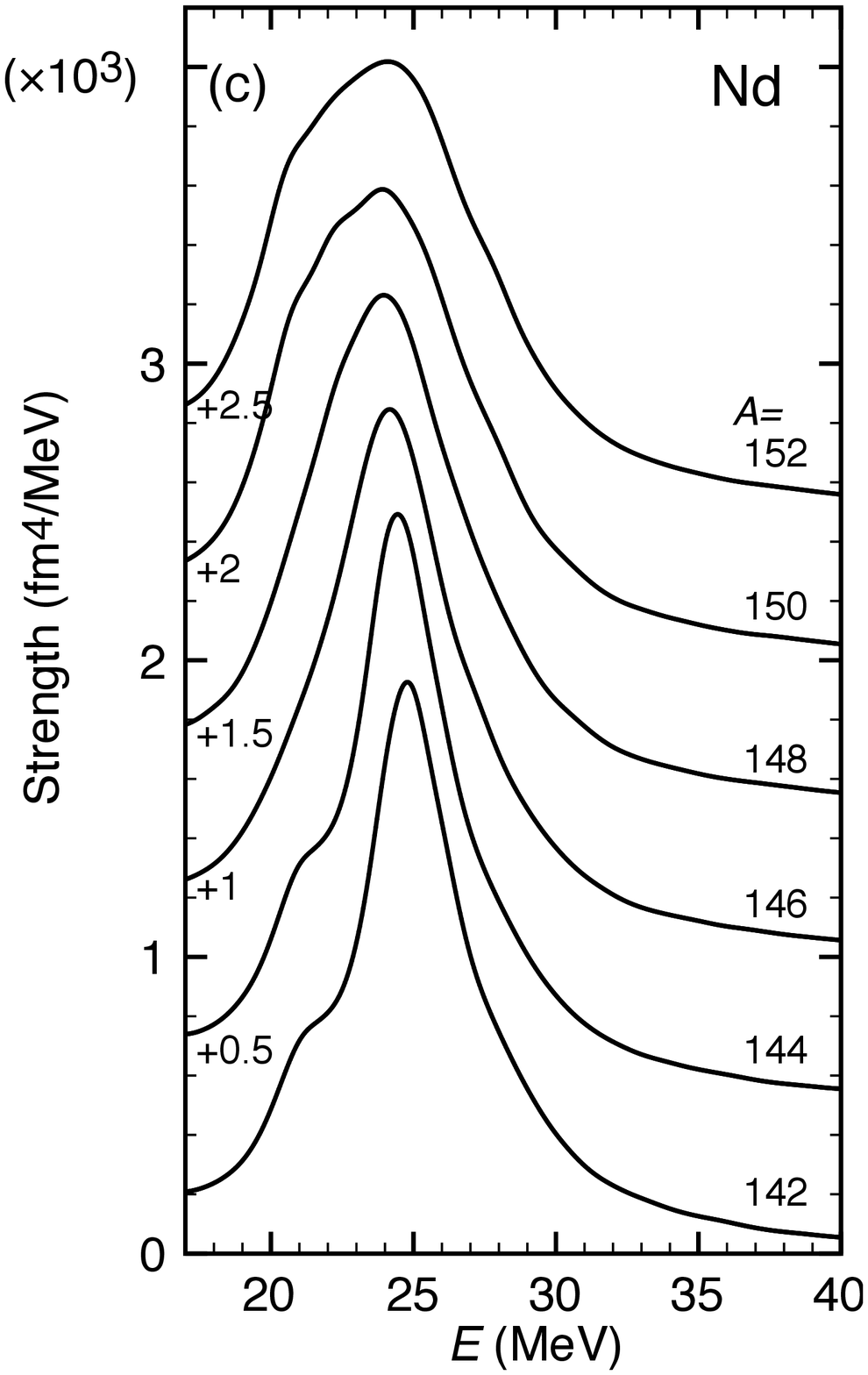}
\includegraphics[scale=0.33]{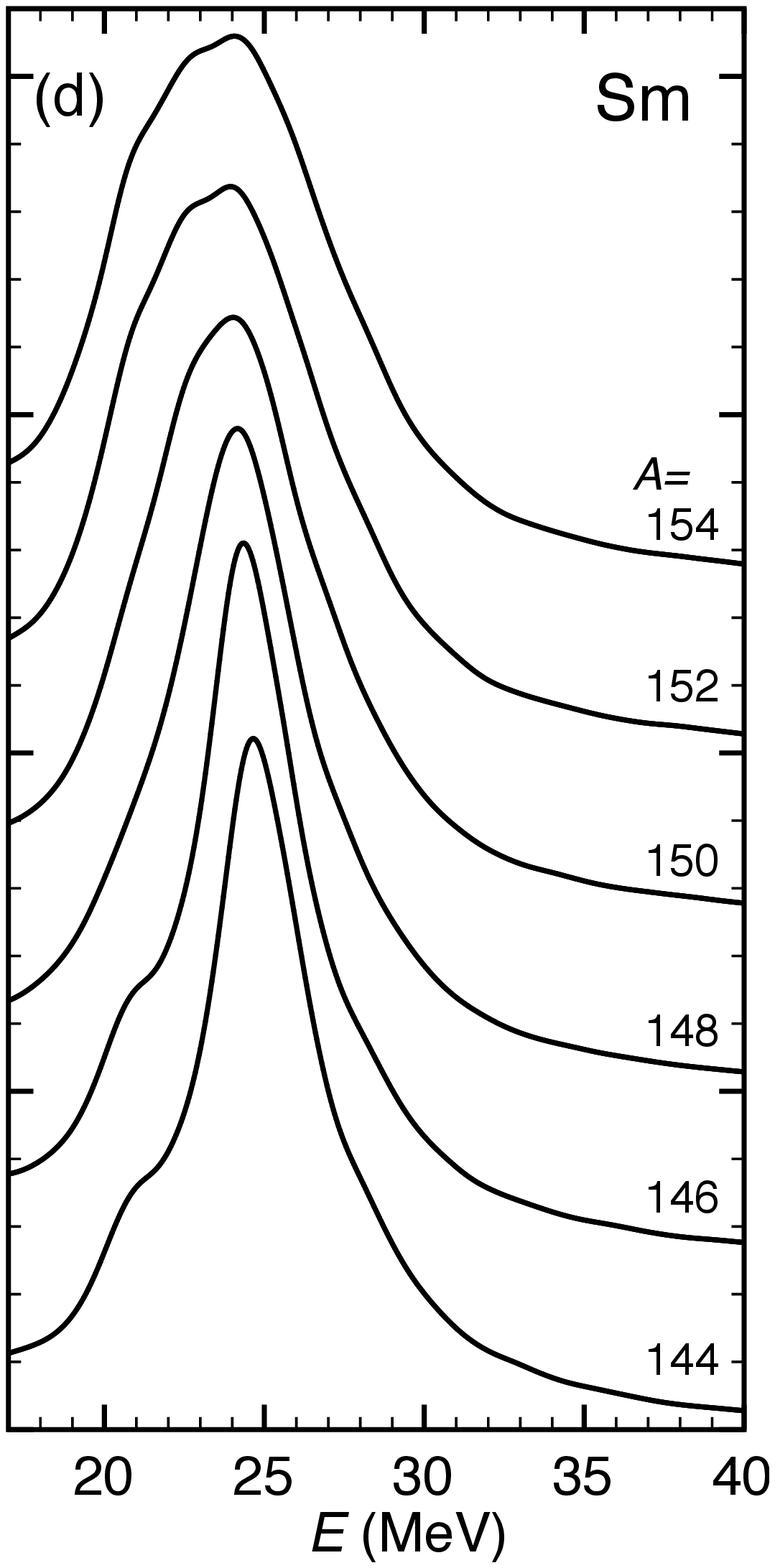}
\caption{The strength distributions (shifted) of IVGMR [(a), (b)] and IVGQR [(c), (d)] 
in Nd and Sm isotopes.}
\label{IV_positive}
\end{center}
\end{figure}

Figure~\ref{IV_positive} shows the strength distributions for the isovector (IV) 
monopole and quadrupole excitations. 
Although the experimental data for the IVGMR and IVGQR 
are unavailable in the mass region under investigation, 
the present calculation suggests the existence of these GR modes
in the Nd and Sm isotopes. 
The energy of IVGQR is approximately fitted 
by $129.5$ and $130.3$ $\times A^{-1/3}$ MeV 
for Nd and Sm isotopes, respectively. 
This is consistent with the experimental observations $\sim 130 A^{-1/3}$ MeV 
in $A=140 - 240$ nuclei~\cite{hen11}. 
The $K$-splitting of the IVGQR in deformed nuclei is invisible
because the $K$-splitting is small.

A double-peak structure can be seen in deformed nuclei for the IVGMR as well as 
for the ISGMR. 
The lower peak around 20 MeV in the deformed nuclei emerges associated with the 
coupling to the $K^{\pi}=0^{+}$ component of the IVGQR. 
The upper peak around 30 MeV may be identified as a primal IVGMR 
because the resonance peak appears in this energy region in the spherical nuclei. 
Similarly to the ISGMR, the upper peak of the IVGMR is upward-shifted with increasing 
the neutron number. This is due to the stronger coupling between 
the IVGMR and the IVGQR in nuclei with larger deformation.
The energy difference between the upper and lower peaks of the 
IVGQR in $^{154}$Sm approaches about 10 MeV, which is more than twice as large as 
the energy difference seen in the ISGMR.

\subsubsection{Negative-parity excitation}

\begin{figure}[t]
\begin{center}
\includegraphics[scale=0.33]{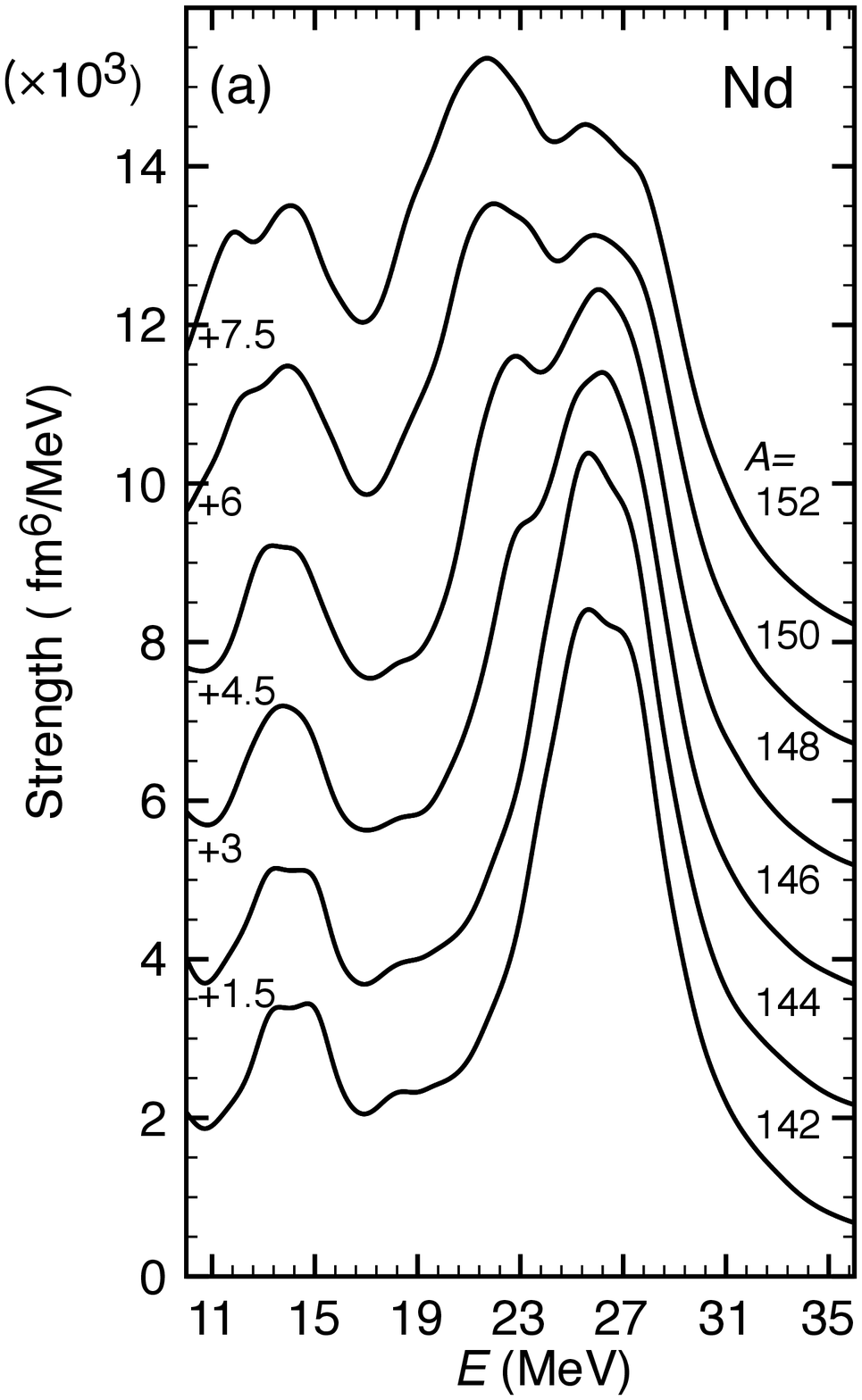}
\includegraphics[scale=0.33]{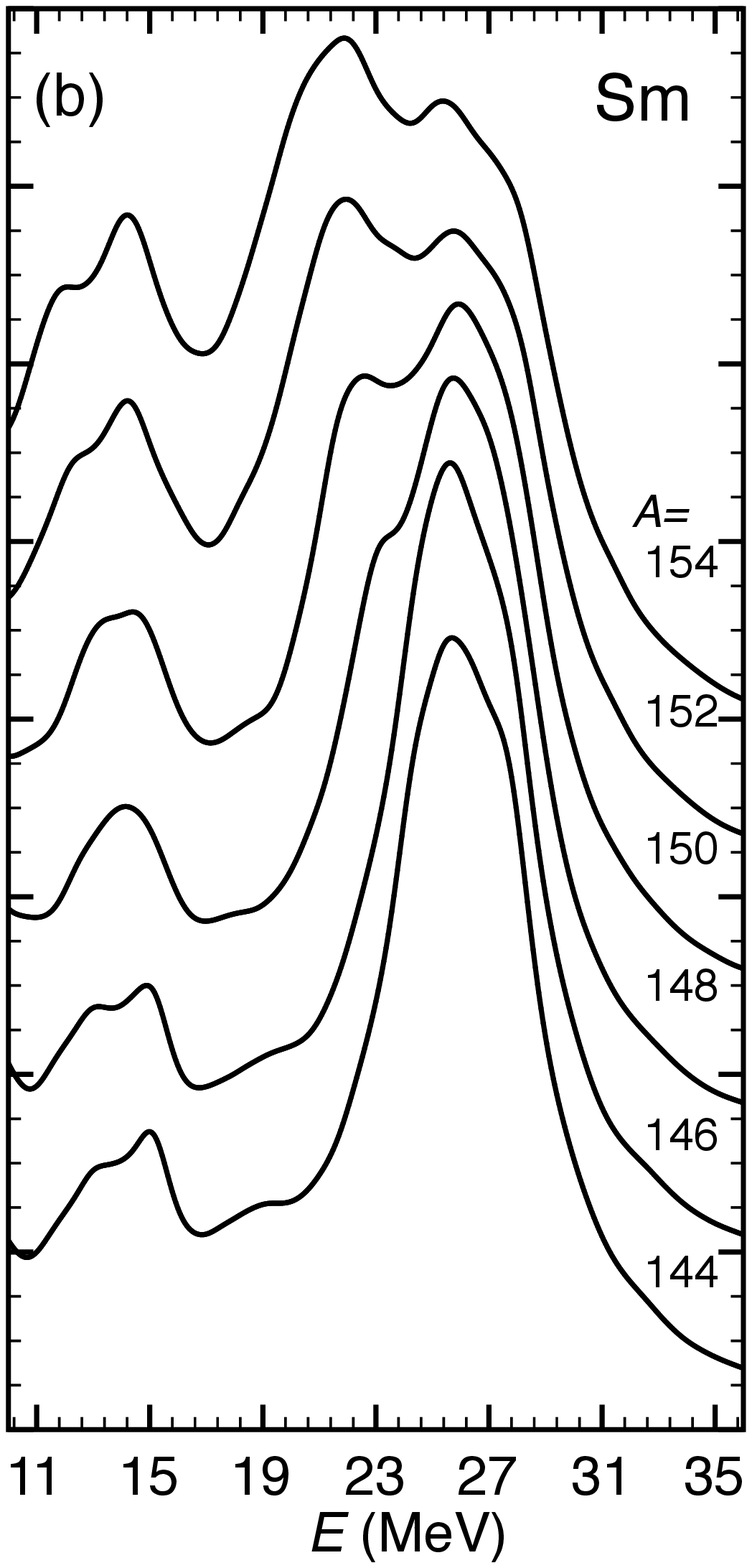} \\
\includegraphics[scale=0.33]{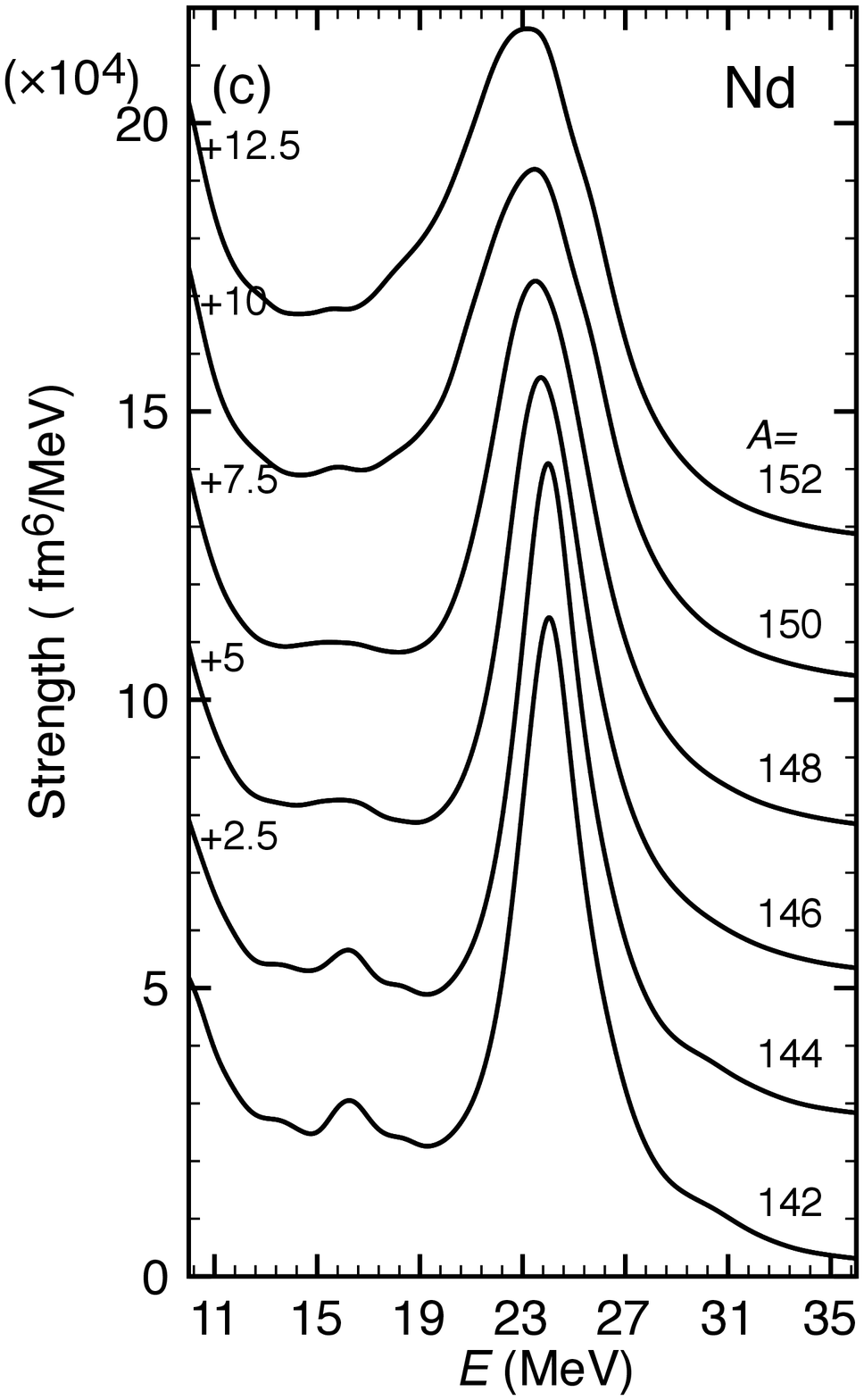}
\includegraphics[scale=0.33]{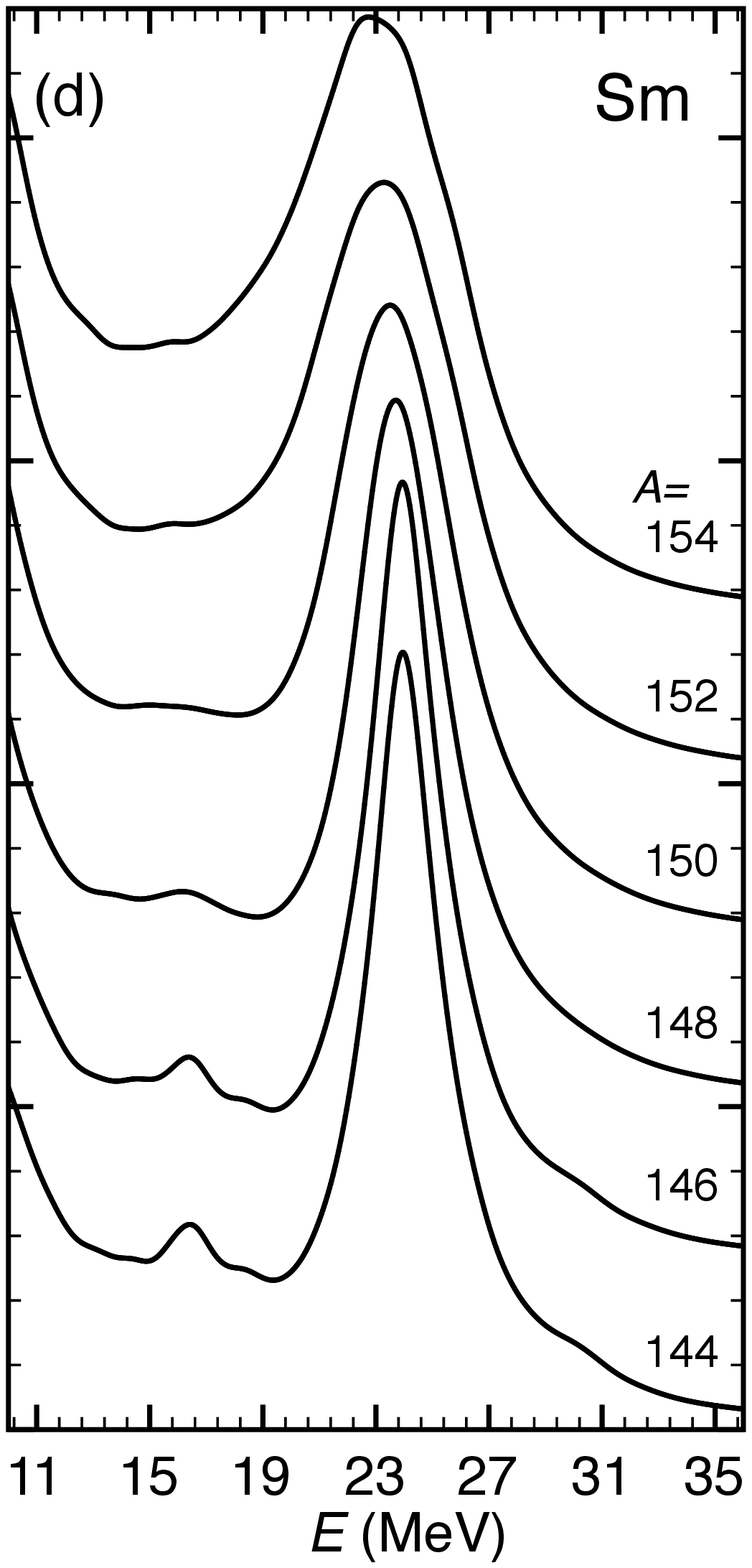}
\caption{Strength distributions (shifted) of ISGDR [(a), (b)] 
and ISGOR (HEOR)  [(c), (d)] in Nd and Sm isotopes.}
\label{IS_negative}
\end{center}
\end{figure}

\begin{figure}[t]
\begin{center}
\includegraphics[scale=0.27]{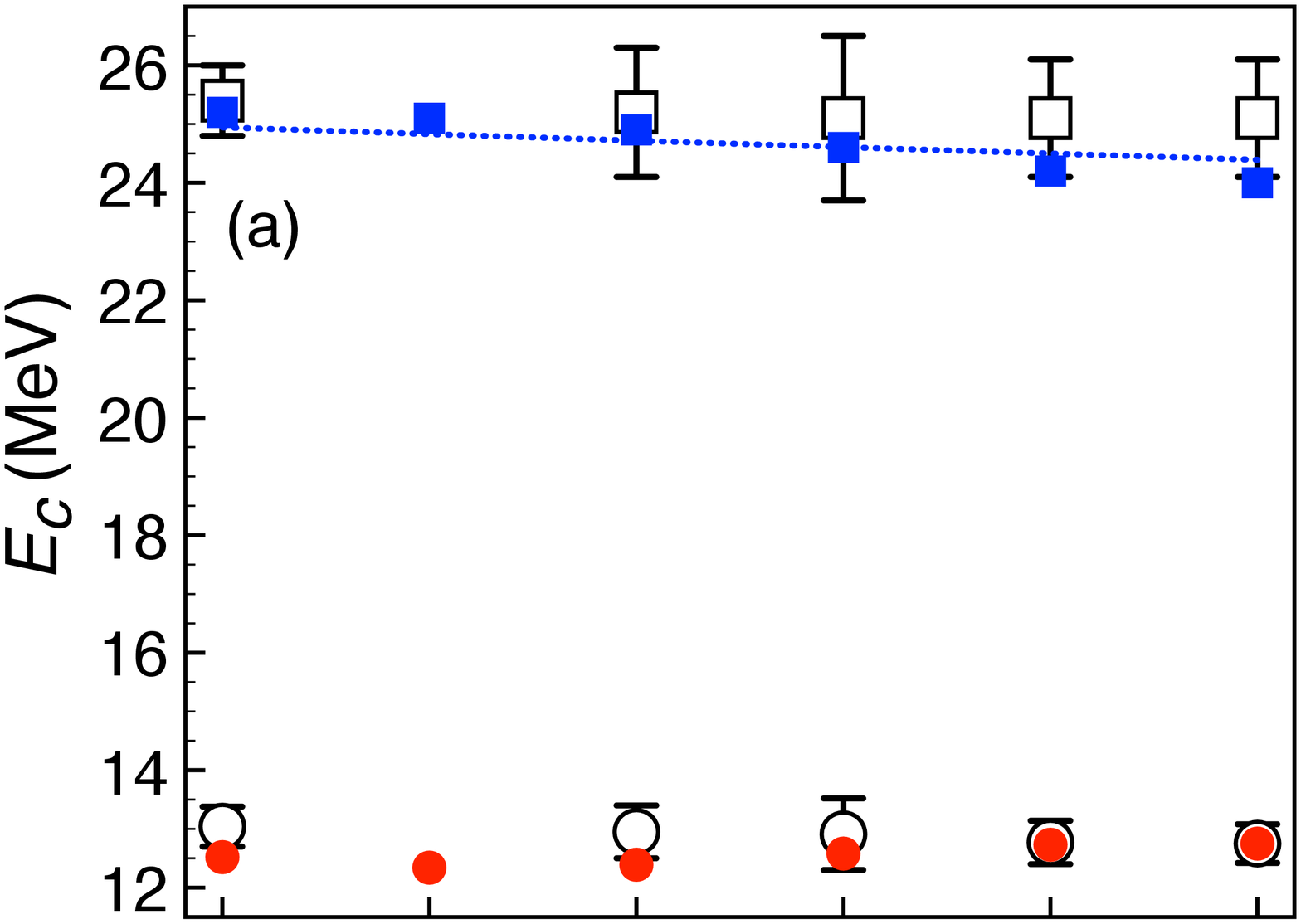}
\includegraphics[scale=0.27]{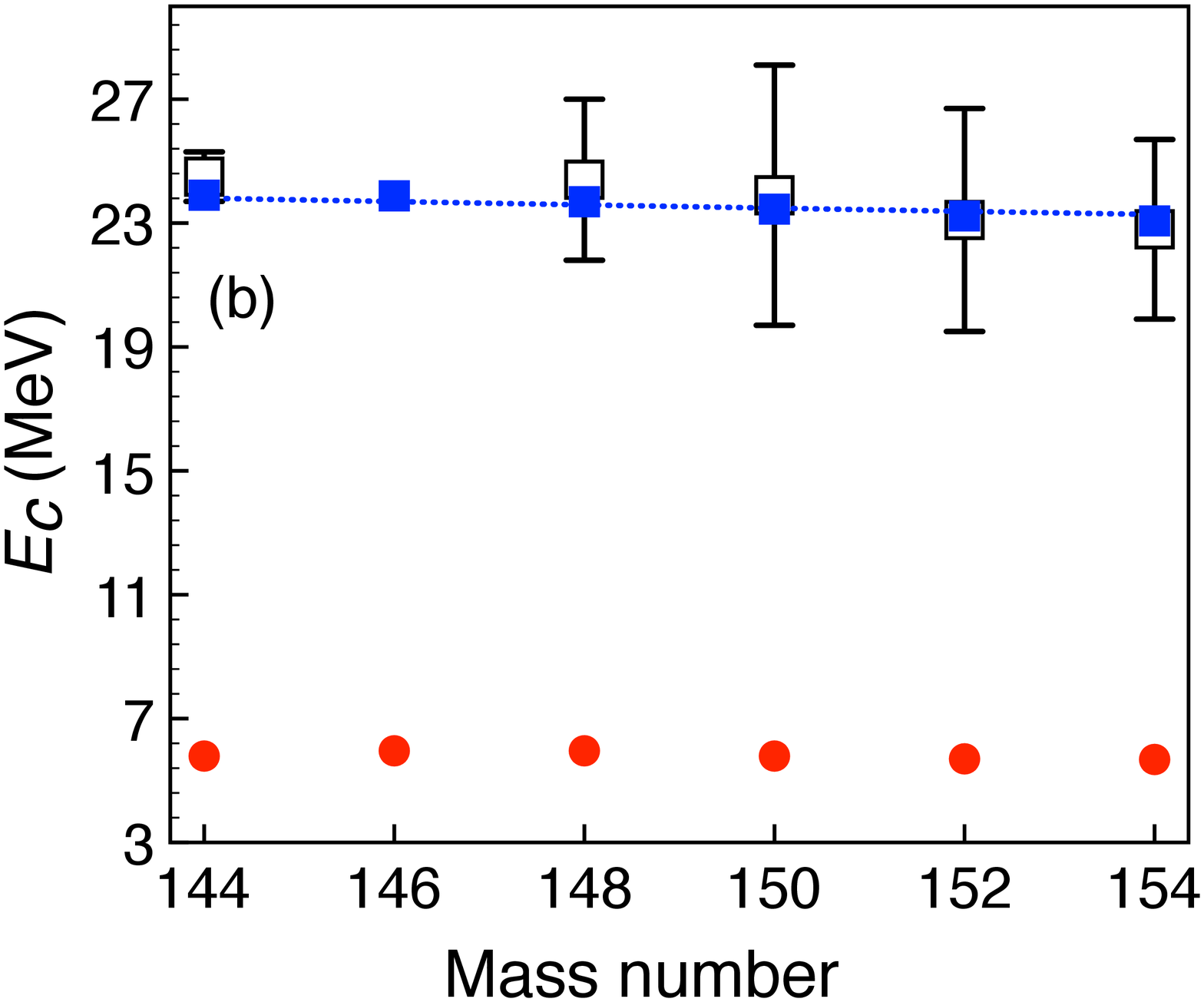}
\caption{(Color online) (a) The centroid energies of the low-energy and high-energy 
components of the ISGDR in the Sm isotopes.  
(b) The centroid energies of the HEOR and the LEOR in the Sm isotopes. 
The centroid energy of LEOR is evaluated in the 
energy range of $[E_a,E_b]=[3,10]$ MeV.
The dotted line is obtained by fitting the results with an $A^{-1/3}$ line. 
The experimental data~\cite{ito03} are denoted by open symbols with error bars.  }
\label{Sm_negative_energy}
\end{center}
\end{figure}

Figure~\ref{IS_negative} shows the strength distributions of 
the IS compression dipole and octupole excitations. 
In the IS octupole-transition-strength distributions, 
we can see a high-energy octupole resonance (HEOR) at around 25 MeV. 
Furthermore, we find a broadening of the width associated with the deformation 
as observed in the experiment~\cite{mor82}. 
We show the centroid energy of the HEOR 
and the low-energy octupole resonance (LEOR) in the Sm isotopes 
in Fig.~\ref{Sm_negative_energy}(b). 
The centroid energy of HEOR and LEOR is evaluated in the 
energy range of [17, 33] MeV and [3,10] MeV, respectively. 
The calculated energy of HEOR is best fitted to a $124.8 \times A^{-1/3}$ line, 
and agrees with the experimental observation~\cite{ito03}.  
However, this excitation energy is 
significantly higher than the systematic value 
of $110 \pm 5  \times A^{-1/3}$ MeV~\cite{har01}.

\begin{figure}[t]
\begin{center}
\includegraphics[scale=0.48]{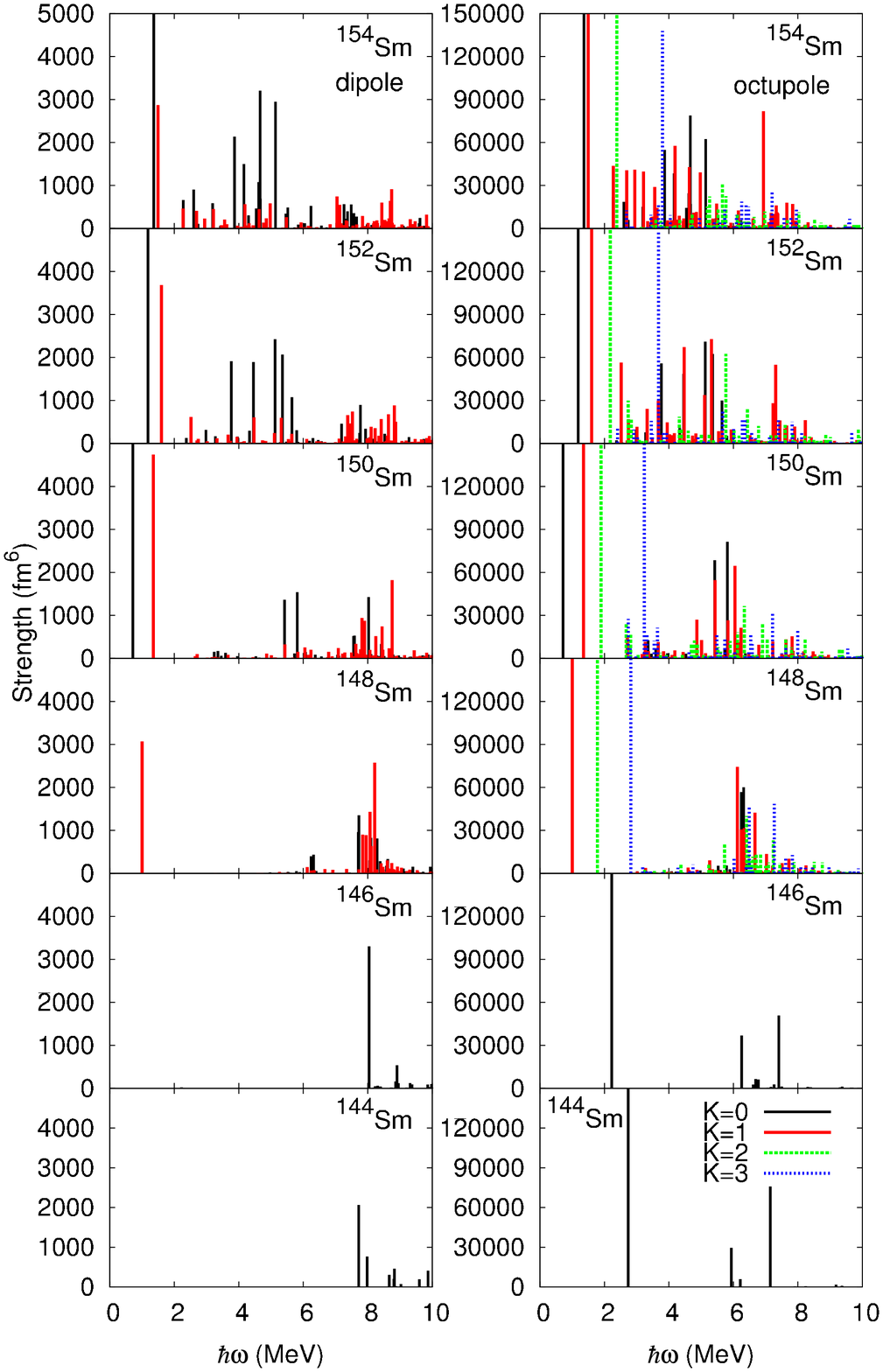}
\caption{(Color online) 
The low-energy IS dipole and octupole transition strengths 
in the Sm isotopes. 
The strengths with different $K$ are all identical for the spherical nuclei.
}
\label{Sm_dip_oct}
\end{center}
\end{figure}

Below 10 MeV, we find low-lying collective (discrete) states and  the LEOR. 
The right panels of Fig.~\ref{Sm_dip_oct} show 
the low-energy part of the IS octupole transition-strengths in the Sm isotopes. 
We find that the low-lying collective $K=2, 3$ states are overlapping with the 
LEOR in the well-deformed nuclei. 
The present calculation gives 6.5$\%$ and 24$\%$ 
of the IS octupole EWSR value in $^{154}$Sm 
for the energy intervals $0-3$ MeV and $0-7$ MeV, respectively.
This is compatible to the experimental value of 
7$\%$ and 19$\%$ for the discrete states only 
and for the low-lying states 
including the discrete levels and the LEOR, respectively~\cite{mos76}. 
The early theoretical calculation 
employing the pairing plus octupole interaction model 
gives also an excellent agreement with the observed value 
by adjusting the interaction strengths~\cite{mal77}.

The calculated octupole strength carries $51-53$ \% of the EWSR value 
in the HEOR energy region of $17-33$ MeV. 
On the contrary, the experiment~\cite{ito03} has reported
decrease of the strength in the same energy region 
from 75\% to 30\% of EWSR as increasing the mass number
in the Sm isotopes.
This inconsistency may be attributed to the uncertainty of the choice 
of the continuum in the experimental analysis 
and the strong overlap with the ISGDR~\cite{har01}.

\begin{figure}[t]
\begin{center}
\includegraphics[scale=0.2]{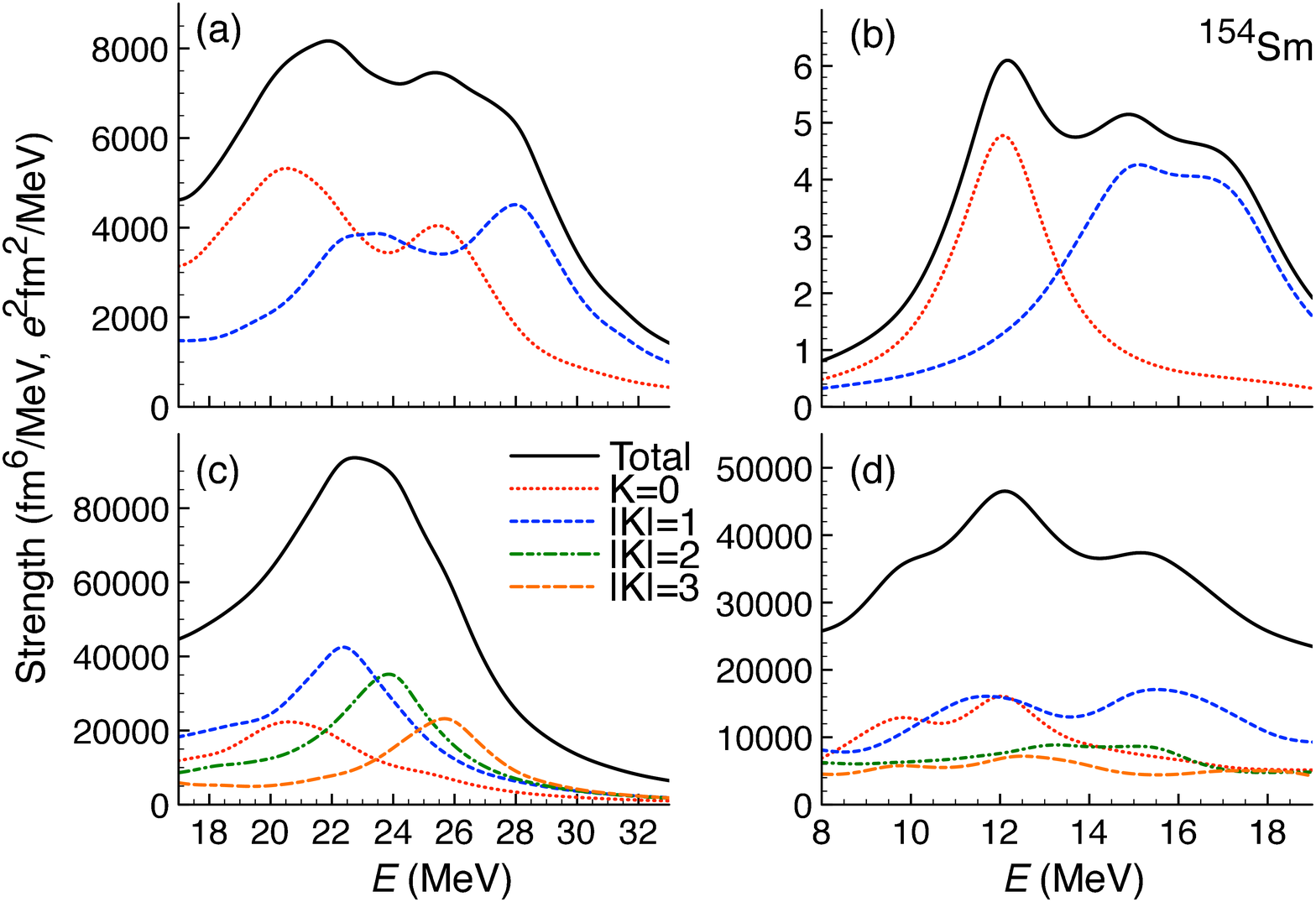}
\caption{(Color online) 
(a) The IS compression dipole strength distribution in the ISGDR energy region in $^{154}$Sm. 
(b) The IV dipole strength distribution in the GDR energy region.
(c) The IS octupole strength distribution.
(d) The IV octupole strength distribution.
}
\label{Sm_ISGDR_HEOR}
\end{center}
\end{figure}

\begin{figure}[t]
\begin{center}
\includegraphics[scale=0.33]{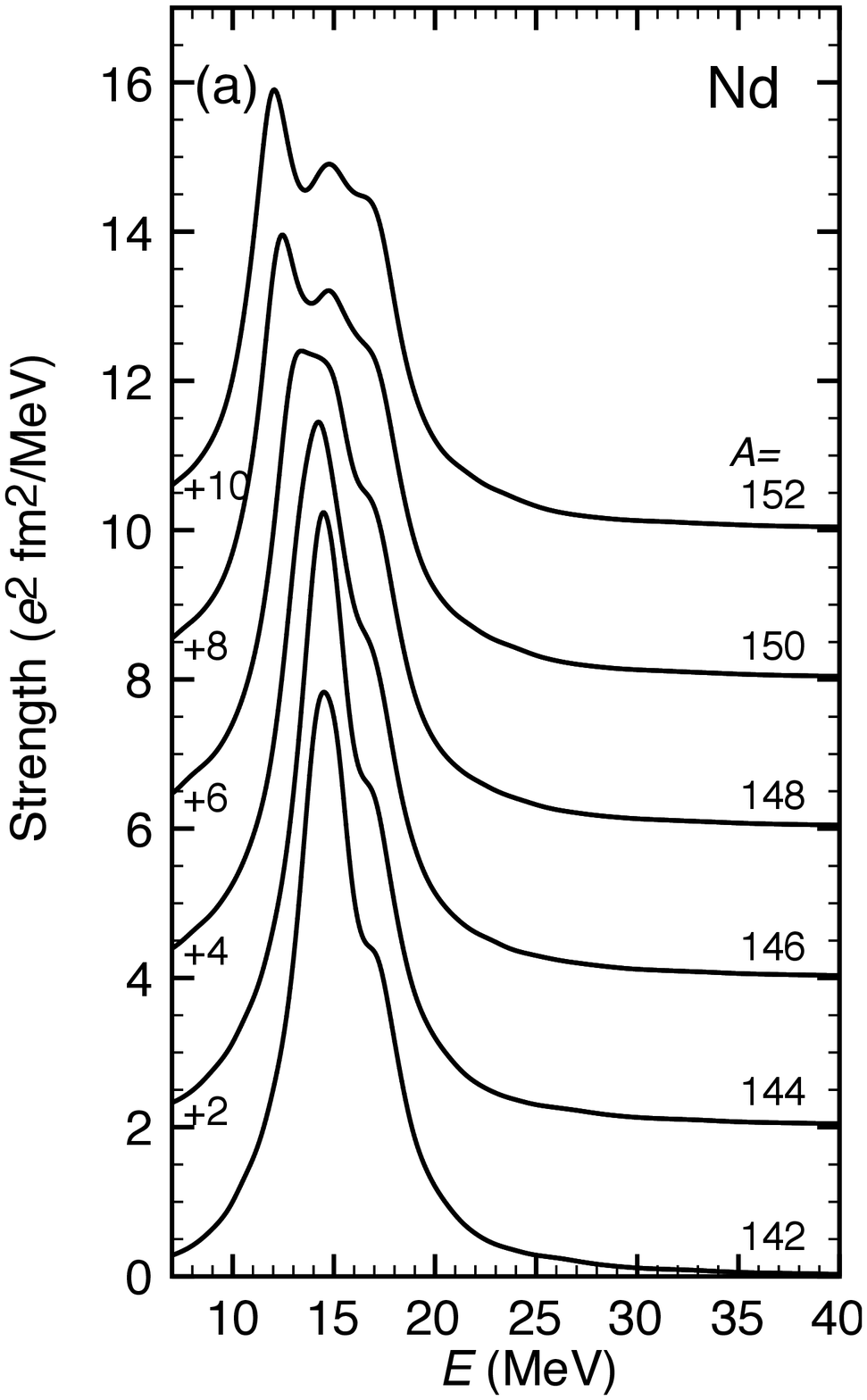}
\includegraphics[scale=0.33]{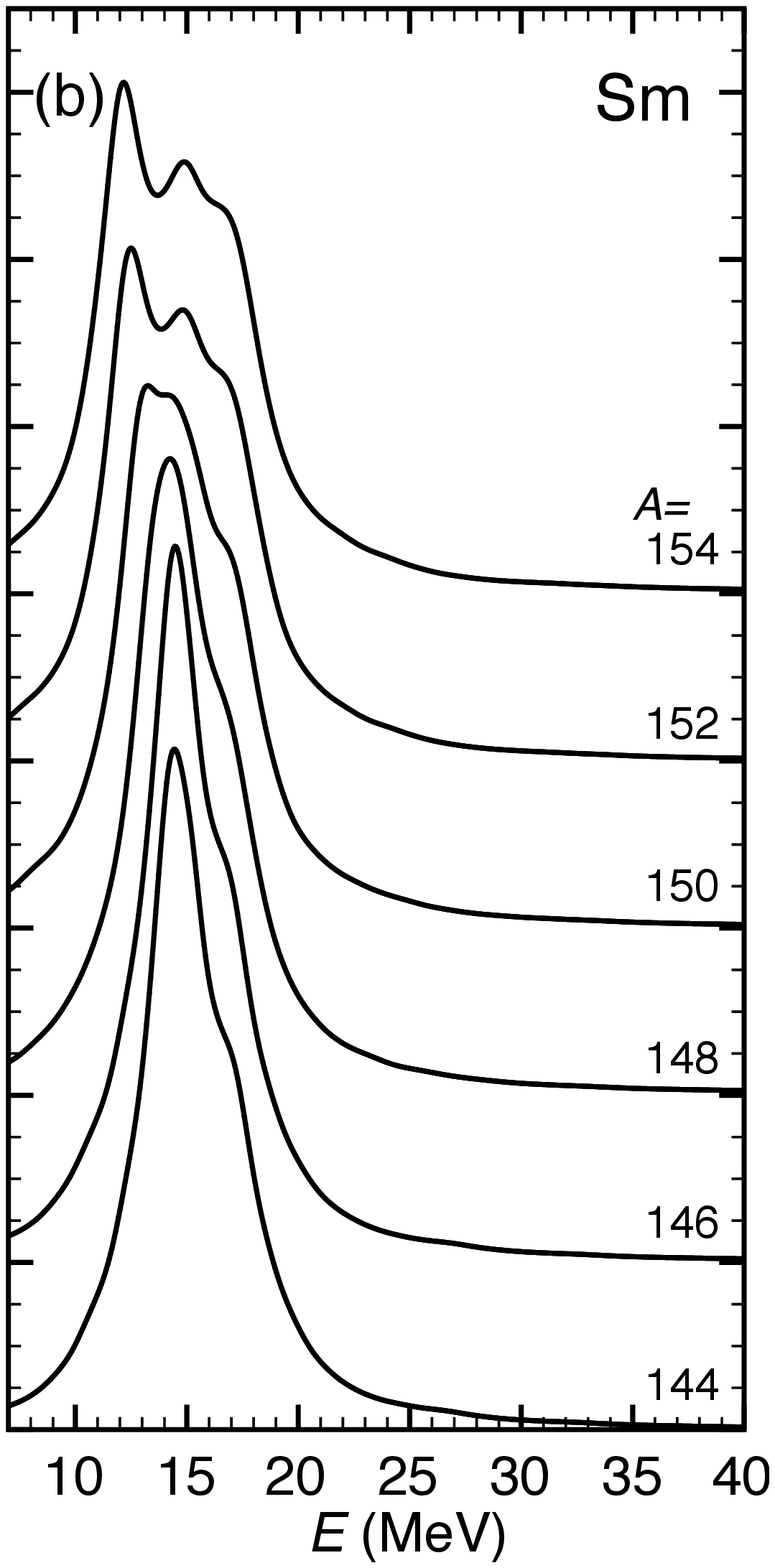} \\
\includegraphics[scale=0.33]{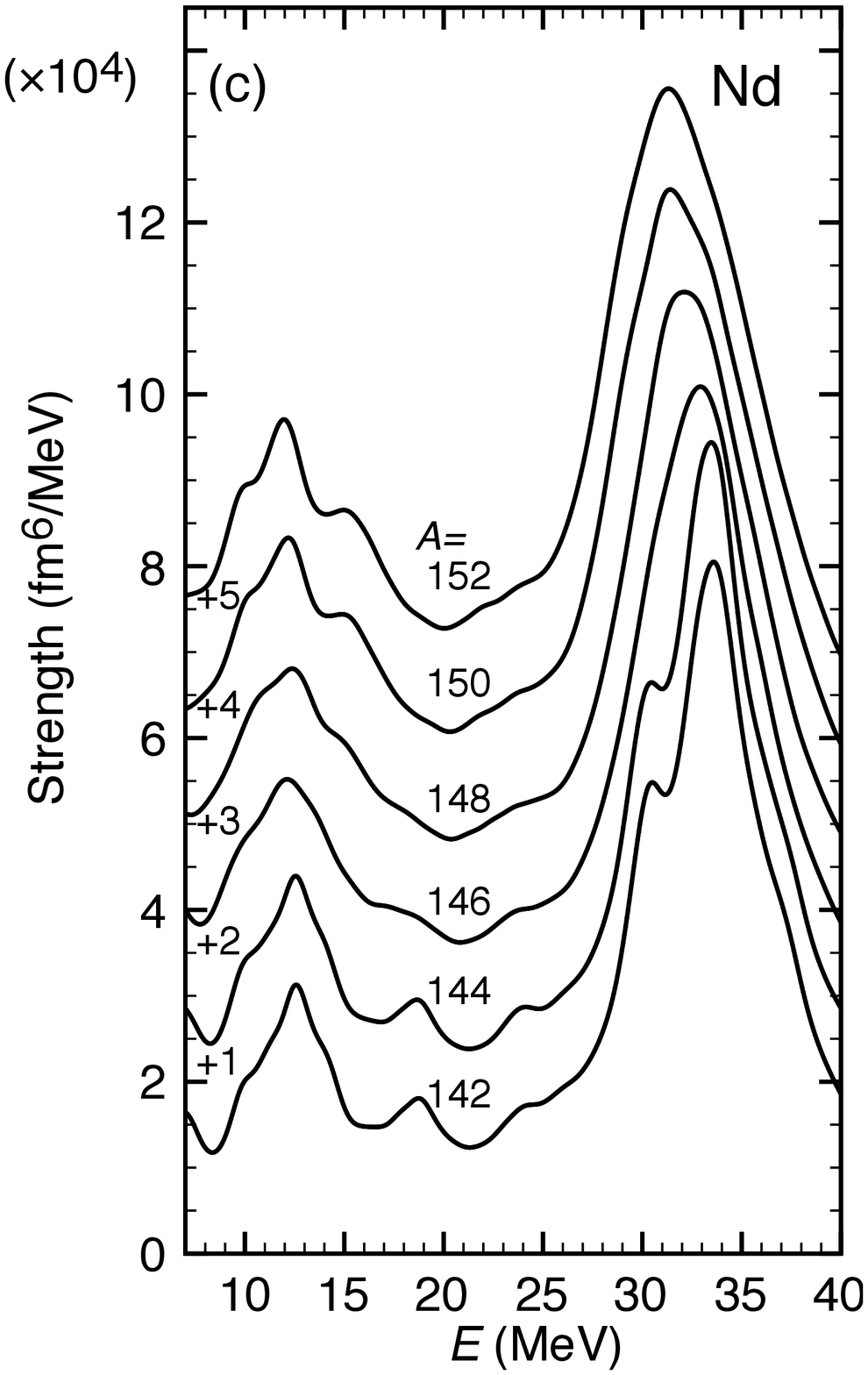}
\includegraphics[scale=0.33]{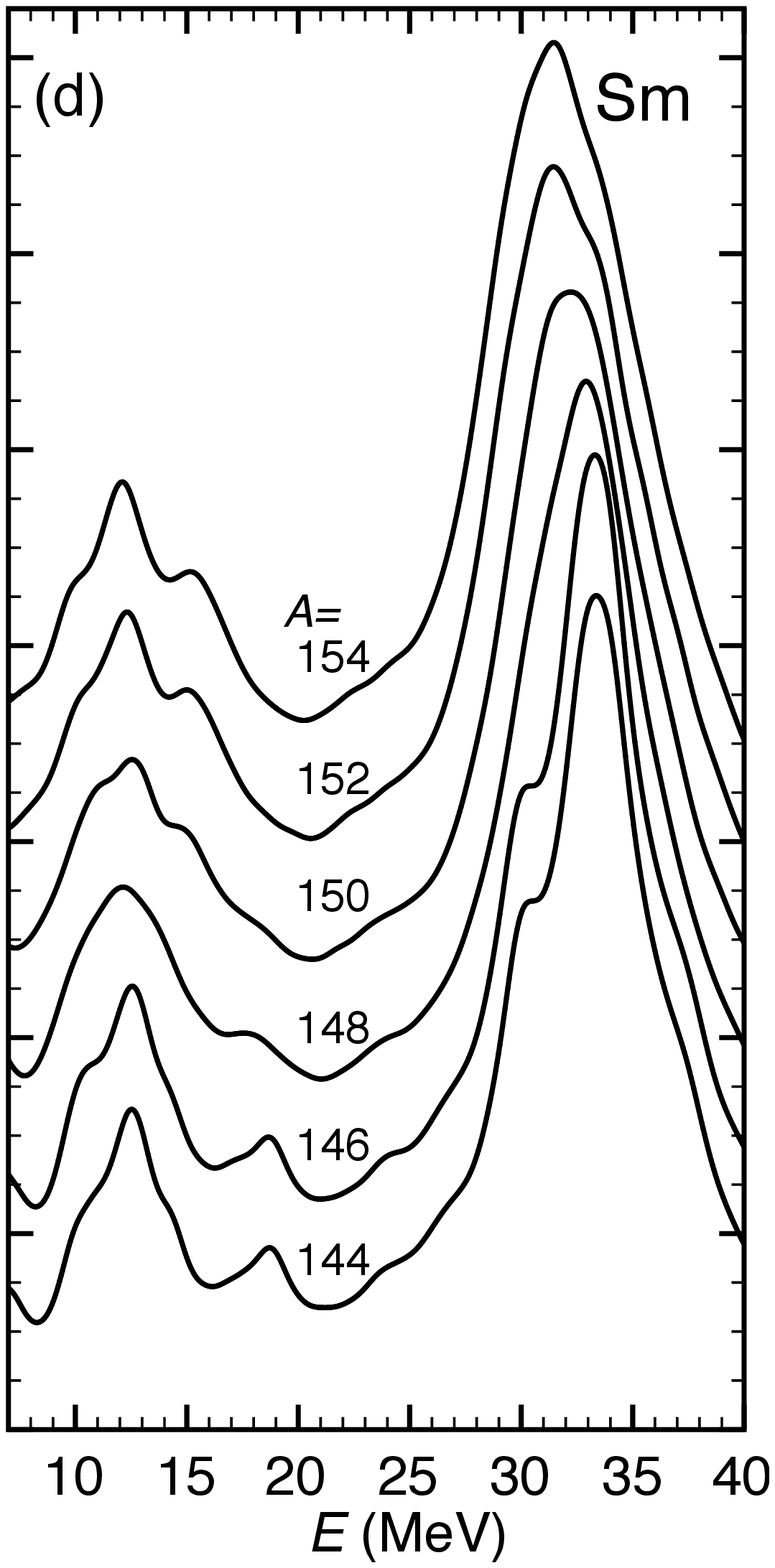}
\caption{The strength distributions (shifted) of IVGDR [(a), (b)] 
and IVGOR [(c), (d)]in Nd and Sm isotopes.}
\label{IV_negative}
\end{center}
\end{figure}

We have the ISGDR at around 25 MeV 
corresponding to the $3\hbar \omega$ excitation, 
and this energy region is where the HEOR is located.  
We show the centroid energy of the ISGDR in the Sm isotopes 
in Fig.~\ref{Sm_negative_energy}(a). 
The calculated energy is best fitted to the $130.7 \times A^{-1/3}$ line. 
The fitted energy of the ISGDR is slightly higher than that of the HEOR. 
The ISGDR in spherical nuclei is investigated 
in the framework of the HF-BCS + QRPA approach employing several Skyrme 
functionals~\cite{col00}. 
The excitation energy obtained in Ref.~\cite{col00} in $^{144}$Sm 
is consistent with our result. 

A deformation effect on the ISGDR can be seen in the increase of its width. 
This is due to the deformation splitting of the $K=0$ and 1 components 
of the ISGDR similarly in the photoabsorption cross sections. 
Furthermore, the width becomes even larger due to 
the coupling to the $K=0$ and 1 components of the HEOR. 
Figures~\ref{Sm_ISGDR_HEOR}(a) and (c) show  the strength distributions 
of the IS dipole and octupole excitations in $^{154}$Sm.  
The resonance structure at $26-28$ MeV appears due to the 
deformation splitting of the primal ISGDR, 
and the structure at $20-23$ MeV is due to the coupling to 
the $K=0, 1$ components of the HEOR. 
Because of these two effects, the total strength 
distribution becomes very broad. 
When we fit the calculated strength distribution with a 
Lorentz line in the energy region of [15, 35] MeV, 
we obtain the width $\Gamma=13.4$ MeV. 
The large width is observed experimentally
as $22.6 \pm 4.2$ MeV in Ref.~\cite{ito03}, 
while the rather small width ($11.8 \pm 0.5$ MeV) is 
reported in Ref.~\cite{you04}.

We furthermore find a low-energy (LE) ISGDR at about 14 MeV. 
We also find that the low-lying dipole states 
appear below 5 MeV with possession of large transition strengths 
in the deformed systems 
as shown in the left panels of Fig.~\ref{Sm_dip_oct}. 
This is due to the coupling to the low-lying octupole modes of excitation. 

The strength distribution in $^{154}$Sm 
obtained by the $(\alpha, \alpha^{\prime})$ experiment in Ref.~\cite{you04} 
shows a three-peak structure at around the excitation energy of $12-16$ MeV, 
$20-24$ MeV and $26-29$ MeV. 
The data were compared with the 
fluiddynamics results of Ref.~\cite{nis85}, however,
the mechanism for appearance of the second peak was unclear.
According to the present calculation, it is suggested that
the first peak corresponds to the low-energy ISGDR,
the second peak is associated with 
the coupling to the $K=0$ and $1$ components of the HEOR, and 
the third peak is the primal ISGDR.

Figure~\ref{IV_negative} shows the strength distributions of 
IV dipole and octupole excitations. 
The IV giant octupole resonance (GOR) is seen above 30 MeV, 
and we find a bump structure at around 10 MeV
corresponding to the IV-LEOR. 
The strength is rather smaller than that of the IV-HEOR. 
Noted that the strength of the IS-LEOR is compatible to that of the IS-HEOR.

In the deformed systems, we see an appearance of 
the shoulder structure at about 15 MeV.
Figure~\ref{Sm_ISGDR_HEOR} (b) and (d) presenting the IV dipole and octupole strength 
distributions in $^{154}$Sm show that the shoulder structure is associated 
with the deformation splitting of the GDR and its coupling to the IV-LEOR.

\subsubsection{Low-lying collective states}

\begin{figure}[t]
\begin{center}
\includegraphics[scale=0.75]{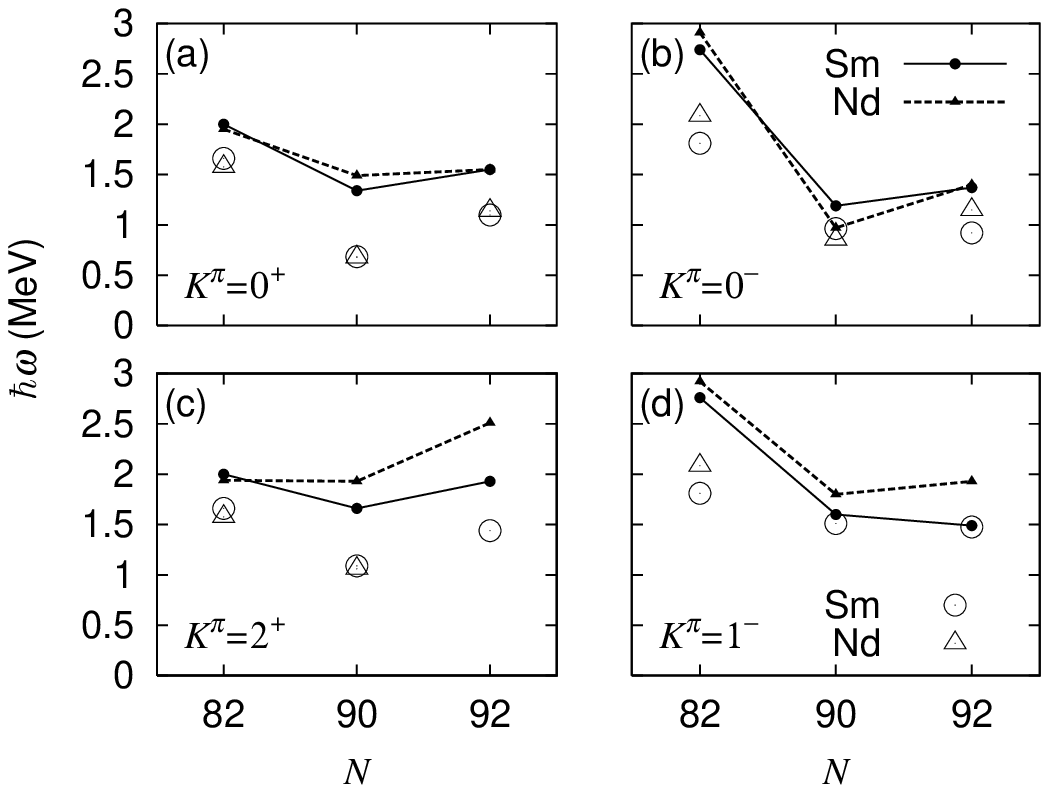}
\caption{
The excitation energies of the lowest $K^{\pi}=0^{+}$, $2^{+}$, $0^{-}$ and $1^{-}$ states 
in Nd and Sm isotopes. The experimental data for Nd and Sm isotopes~\cite{ENSDF} are denoted 
by open triangles and circles, respectively. 
The lines are drawn to guide the eyes.
}
\label{Sm_Nd_energy}
\end{center}
\end{figure}

In this subsection, we are going to discuss 
the low-lying states.
As shown in Fig.~\ref{Sm_dip_oct}, 
we see an appearance of the 
collective mode for the IS dipole excitation below 2 MeV 
associated with an onset of deformation. This is due to the strong coupling to 
the collective octupole mode of excitation. 

What has to be mentioned here is an absence of the collective $K=0$ mode 
in $^{148}$Sm. In the present calculation, we have two imaginary solutions 
in the $K^{\pi}=0^-$ channel, one of which is associated with the spurious 
center-of-mass motion. 
In $^{150}$Sm, we have the $K=0$ mode at 0.72 MeV. 
The excitation energy of the collective $K=0$ mode becomes higher 
when increasing the neutron number. 
Thus, we can consider that the second imaginary solution
in $^{148}$Sm indicates 
the instability against the axially-symmetric octupole deformation. 
In fact, the largest $B(E3; 0_1^+ \to 3_1^-)$ value 
is measured in $^{148}$Sm among the even-even Sm isotopes~\cite{ENSDF}.

Before going to the next subsection, 
we summarize the energy of the low-lying collective states in the spherical 
and the well-deformed Nd and Sm isotopes. 
Figure~\ref{Sm_Nd_energy} shows the excitation energies of the lowest 
$K^{\pi}=0^{+}$, $2^{+}$, $0^{-}$ and $1^{-}$ states. 
The available experimental data~\cite{ENSDF} are also shown. 
For the experimental values, 
we neglect the rotational correction, 
which is $30$ keV at most in $^{154}$Sm. 
Figure~\ref{Sm_Nd_energy} shows that 
the observed isotopic dependence is well reproduced.

The excitation energies of the quadrupole-vibrational states 
agree with the experimental data within $0.5 - 1$ MeV. 
This result is close to the one obtained in Ref.~\cite{ter11}, 
where they obtained the $\gamma$-vibrational state at 2.5 MeV and at 2.3 MeV 
in $^{152}$Nd and in $^{154}$Sm, respectively 
despite the use of a different pairing functional from ours.
Reproduction of the experimental values of the octupole-vibrational 
states in the deformed nuclei is extremely good.

\begin{table}[t]
\begin{center}
\caption{The excitation energies (in units of MeV) of the low-lying collective states in $^{154}$Sm. 
Experimental data are taken from Ref.~\cite{ENSDF}.}
\label{154Sm_lowlying}
\begin{tabular}{ccccc}
\hline \hline
\noalign{\smallskip}
 & $K^{\pi}=0^{+}$ & $K^{\pi}=2^{+}$ & $K^{\pi}=0^{-}$ & $K^{\pi}=1^{-}$  \\
\noalign{\smallskip}\hline\noalign{\smallskip}
SkM* & 1.55 & 1.93 & 1.37 & 1.49  \\
SLy4 & 1.46 & 1.81 & 1.25 & 1.66  \\
SkP & 0.95 & 0.92 & 1.44 & 1.64  \\
Exp. & 1.099 & 1.440 & 0.921 & 1.475 \\
\noalign{\smallskip}
\hline \hline
\end{tabular}
\end{center}
\end{table}

Table~\ref{154Sm_lowlying} summarizes the excitation energy of the low-lying 
collective states in $^{154}$Sm obtained by the QRPA calculations 
employing the different kinds of Skyrme functionals. 
All the Skyrme functionals under consideration give a reasonable agreement 
with the measurements, and the quality is at the same level found in Ref.~\cite{ter10}.

\subsection{Incompressibility and effective mass in GRs}

\begin{figure}[t]
\begin{center}
\includegraphics[scale=0.25]{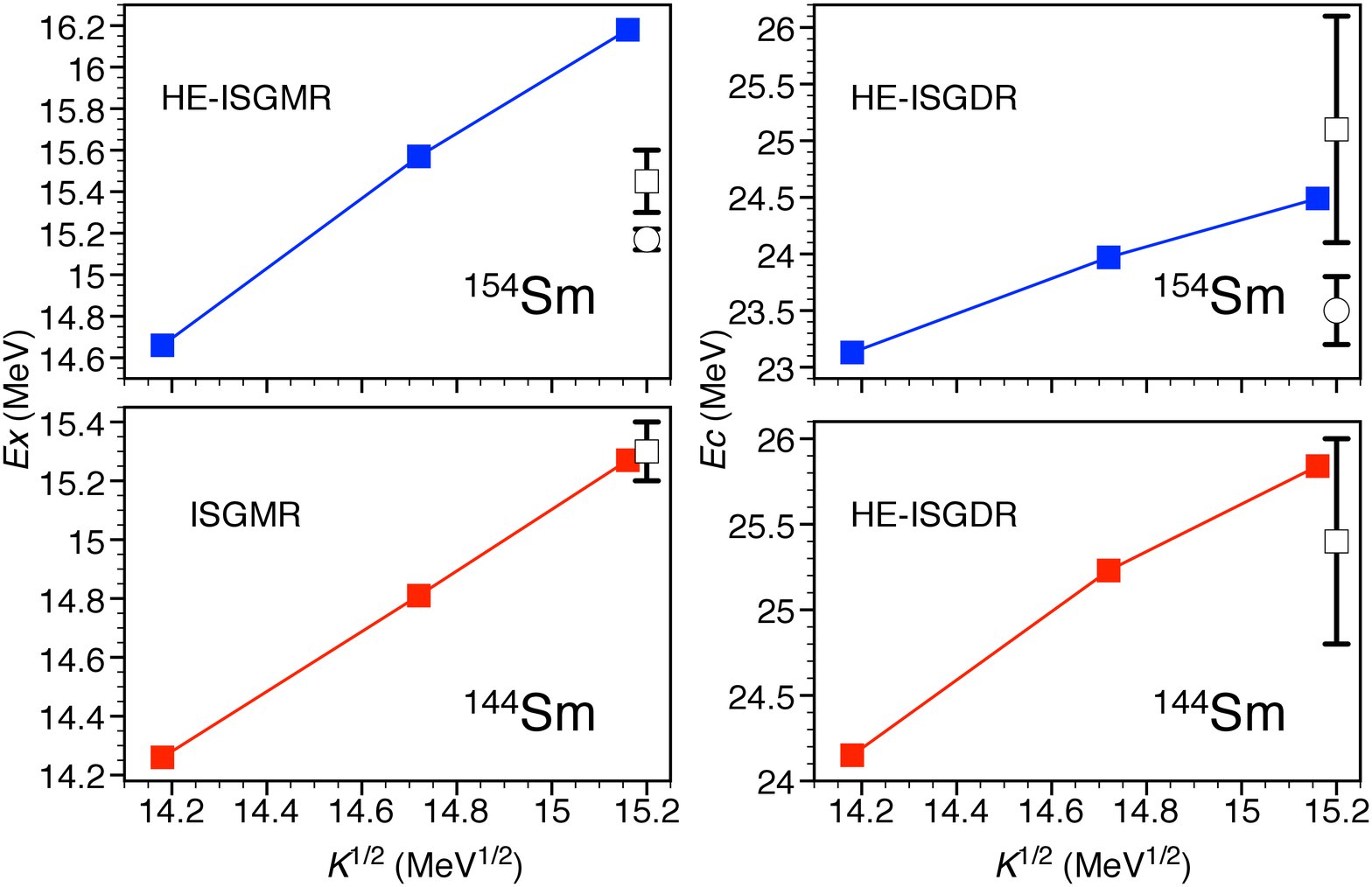}
\includegraphics[scale=0.25]{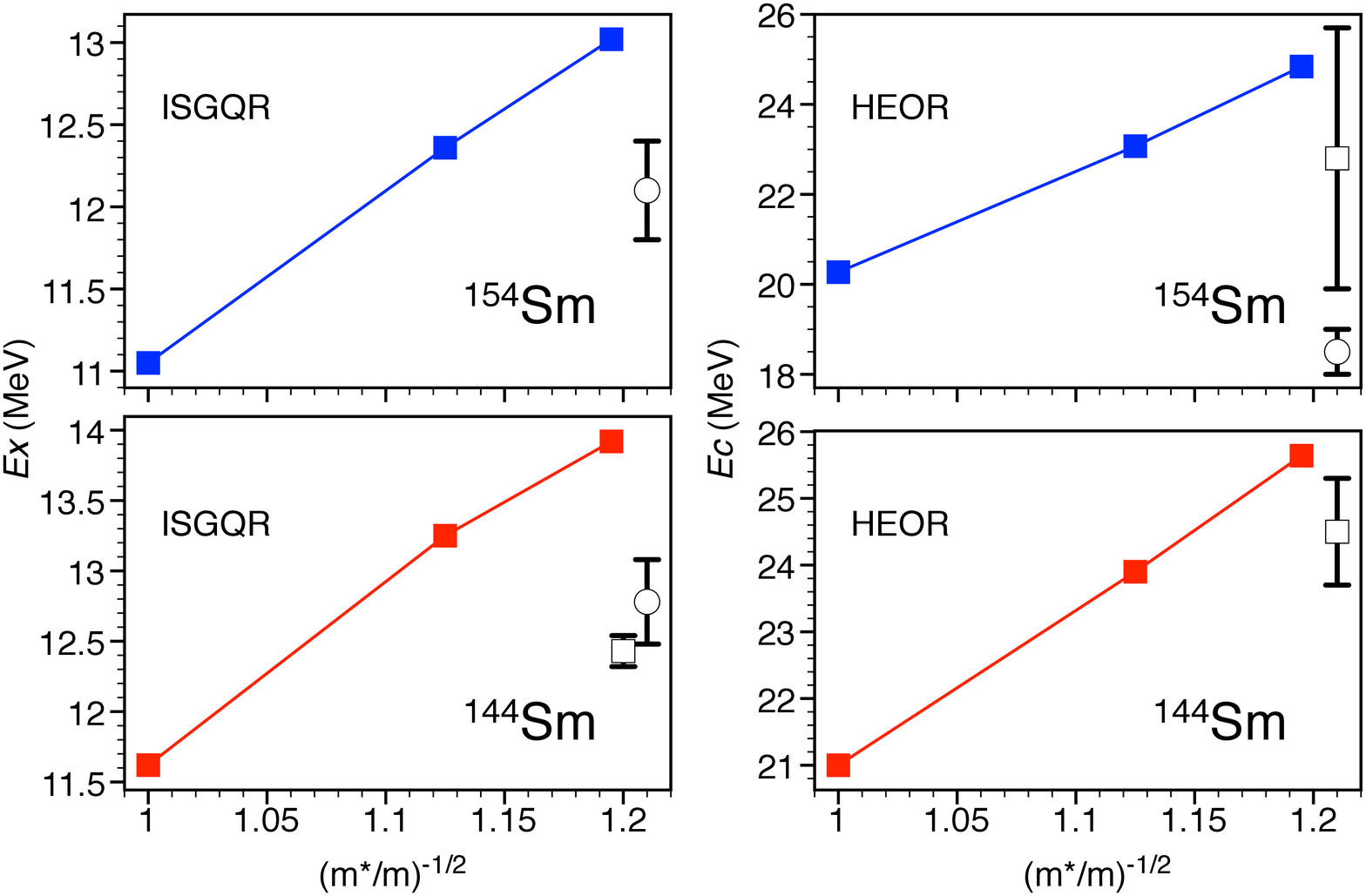}
\caption{(Color online) The peak energies of the ISGMR and ISGQR,
and the centroid energy of the ISGDR and HEOR in $^{144}$Sm and $^{154}$Sm 
obtained by employing the Skyrme functionals giving the different nuclear-matter properties. 
The continuous lines are drawn to guide the eyes. 
Experimental data are shown by the open square~\cite{ito03} and the circle~\cite{you04} 
with the error bars.}
\label{Sm_GMR_GQR_HEOR_energy}
\end{center}
\end{figure}

In this subsection, we investigate 
how the calculated properties of 
the GRs depend on the Skyrme EDFs
with the different nuclear matter properties, effective mass and the incompressibility. 
We take $^{144}$Sm and $^{154}$Sm
as examples of spherical and deformed nuclei, respectively.
The experimental data for all the isoscalar multipole excitations
are available for these isotopes.
Nuclear matter and deformation properties for 
the functionals we employ are listed in Table~\ref{Skyrme}. 

\begin{table}[t]
\begin{center}
\caption{The nuclear matter and deformation properties for the Skyrme functionals 
under consideration. The table represents the isoscalar effective mass $m^{*}_{0}/m$, 
the nuclear-matter incompressibility $K$, 
and the deformation parameters $\beta_{2}$ in $^{154}$Sm.}
\label{Skyrme}
\begin{tabular}{ccccc}
\hline \hline
\noalign{\smallskip}
Forces  & $m^{*}_{0}/m$ &  $K$ (MeV) & $\beta_{2}^{\nu}$ & $\beta_{2}^{\pi}$ \\
\noalign{\smallskip}\hline\noalign{\smallskip}
SkM* & 0.79  & 216.7 & 0.30 & 0.33 \\
SLy4 & 0.70  & 229.9 & 0.30 & 0.33 \\
SkP & 1.00 & 201.0 & 0.28 & 0.30 \\
\noalign{\smallskip}
\hline \hline
\end{tabular}
\end{center}
\end{table}

As we discussed in Section~\ref{positive}, 
the experimental value for the peak energies of the ISGMR 
is fairly reproduced in the calculation for all the functionals under investigation. 
The excitation energies of the upper peak of the ISGMR in $^{154}$Sm 
and the ISGMR in $^{144}$Sm are shown 
in the upper-left panels of Fig.~\ref{Sm_GMR_GQR_HEOR_energy} 
as functions of the square root of the incompressibility. 
We can see a clear correlation between them. 
This result is consistent with the fact that the GMR energy is 
proportional to the square root of the incompressibility~\cite{bla80}. 
The excitation energy is given in the scaling model as
\begin{equation}
\omega^{s}_{\mathrm{M}}=\sqrt{5(m/m^{*}_{0})(1+F_{0})}
\Omega=\sqrt{5K/3m\langle r^{2}\rangle},
\end{equation}
where $F_{0}$ is the Landau-Migdal parameter and 
$\hbar\Omega\simeq 41A^{-1/3}$ (MeV)~\cite{nis85}, 
and the excitation energy of the upper peak of the ISGMR in deformed systems 
is given in Eq.~(3.10) of Ref.~\cite{nis85}. 
Note here that as we saw in Sec.~\ref{positive} the scaling model overestimates 
the energy of the compressible modes, 
while it gives the qualitative understanding of GRs~\cite{nis85}.
Since the SkP functional has a small incompressibility, 
the calculated excitation energy of the ISGMR is lower than the experimental data 
and the results obtained by using the SkM* and SLy4 functionals.

For the GMR in $^{154}$Sm, the SkM* functional gives the excitation 
energy which is very close to the observation~\cite{ito03}. 
However, in $^{144}$Sm the SkM* underestimates the observation, 
and the SLy4 gives the reasonable energy. 
The experimental data reported in Refs.~\cite{you99,you04} for the 
GMR centroid energy in $^{144}$Sm 
are $15.39 \pm 0.28$ MeV and $15.40 \pm 0.30$ MeV. 
Therefore, the present calculation suggests that the nuclear-matter 
incompressibility is about $15.2^2 = 231$ MeV deduced from the comparison 
for $^{144}$Sm and $14.6^2=213$ MeV for $^{154}$Sm.
As mentioned in Section~\ref{GS_property}, 
the pairing properties in $^{144}$Sm and $^{154}$Sm 
are quite different. 
Thus it would be interesting to investigate in detail the pairing effects on the GMR~\cite{li08,kha10}, 
taking the deformation effect into account.

The upper-right panel of Fig.~\ref{Sm_GMR_GQR_HEOR_energy} shows the 
the centroid energy of the ISGDR. 
Here, we evaluate the centroid energy 
in the energy region of the second and the third peaks as done in 
the experimental analysis~\cite{ito03,you04} for $^{154}$Sm.
The excitation energy of the ISGDR is given in the scaling model as~\cite{nis85}
\begin{equation}
\omega_{\mathrm{D}}^{s} = \sqrt{ \dfrac{7}{3}
\left[ \dfrac{5K}{3m\langle r^2 \rangle} + \dfrac{8}{5}\left(\dfrac{m}{m^*_0} \right) \right] }\Omega. 
\end{equation} 
It contains information not only of the incompressibility but of the effective mass. 
Note that the primal ISGDR in the deformed nuclei 
is the third peak as we discussed in the previous section.

In the left-lower and right-lower panels of Fig.~\ref{Sm_GMR_GQR_HEOR_energy}, 
we show the peak energy of 
the ISGQR and the centroid energy of the HEOR 
as functions of the inverse of square root of the isoscalar effective mass 
$\sqrt{m^{*}_{0}/m}^{-1}$. 
We can see a linear correlation between them: 
The smaller the isoscalar effective mass, the higher the resonance energy. 
This is consistent with the results of the simple model.
The excitation energy of the ISGQR and HEOR is given by the scaling model as~\cite{nis85}
\begin{align}
\omega^{s}_{\mathrm{Q}}&=\sqrt{2(m/m^{*}_{0})}\Omega, \label{GQR_scaling}\\
\omega^{s}_{\mathrm{O}}&=\sqrt{(28/5)(m/m^{*}_{0})}\Omega. \label{HEOR_scaling}
\end{align}
This feature is also consistent with the finding in the GQR energy obtained 
by the RPA calculations for spherical systems~\cite{rei99}. 

For the ISGQR, the effective mass $m^*_0/m$ around $1.09^{-2}=0.84$ gives the 
excitation energy which is compatible with the experimental results both in 
$^{144}$Sm and in $^{154}$Sm.
For the HEOR in $^{144}$Sm, slightly smaller effective mass 
around $1.16^{-2}=0.74$ seems to be favored 
in comparison with the experimental observation~\cite{ito03}. 
In $^{154}$Sm, it is hard to deduce the optimal value for the 
effective mass due to the large error in the experiment~\cite{ito03}. 

The excitation energies of the HEOR in $^{144}$Sm and $^{154}$Sm 
reported in Ref.~\cite{you04} are $19.6 \pm 0.5$ MeV and $18.5 \pm 0.5$ MeV, respectively. 
The error is much smaller than that in the experiment at RCNP~\cite{ito03}. 
However, the excitation energy is small and it is outside of the energy region 
obtained by the three types of Skyrme functionals. 
This indicates that the effective mass is around $0.95^{-2}=1.11$ or
even larger.
Since the strength distribution of the ISGDR in Ref.~\cite{you04} looks 
similar to our results, the large discrepancy found in the HEOR is 
difficult to understand.

\section{Summary}\label{summary}

We have investigated the deformation effects on GRs 
in the rare-earth nuclei 
by employing the newly developed parallelized 
computer code for the QRPA based on the Skyrme EDF. 
We found a good scalability for the calculation of the matrix elements 
of the QRPA equation by the use of a two-dimensional block 
cyclic distribution, which is suited for the ScaLAPACK.

The axial deformation in the ground state allows the GRs
with the multipolarity $L=0$ and $2$ 
to mix in the $K^{\pi}=0^{+}$ channel.
Accordingly, we have obtained a double-peak structure of the ISGMR. 
The energy difference between the upper and lower peaks
in the ISGMR and the fraction of the energy-weighted summed
strength in the lower peak can be a sensitive measure of
the ground-state quadrupole moment. 
We also predict a prominent double-peak structure of the IVGMR.

For the negative-parity excitations, 
the excitation modes with $L=1$ and $3$ 
can mix in the $K^{\pi}=0^{-}$ and $1^{-}$ channels. 
This mixing leads to a large width for the ISGDR 
and the enhancement of the
low-lying dipole-transition strengths associated with coupling to the 
collective octupole mode of excitation. 
In the IV channel, the excitation energies of GDR and LEOR are similar.
In deformed nuclei, the coupling between these two modes creates a
broadening of the IV-LEOR peak.

It should be emphasized here that the origin of the observed peak splitting
in the IVGDR is different from that of the other GRs.
The double-peak structure in the IVGDR is well-known to be due to
a direct consequence of the nuclear deformation \cite{BM2}.
Namely, this is associated with different frequencies between $K=0$
and $K=1$ modes in the axially deformed system.
The same kind of deformation splitting,
according to the different $K$ quantum numbers, also exists in the
other GRs, however, its magnitude is much smaller than the IVGDR.
Typically, the magnitude of the $K$-splitting is about 2 MeV.
Therefore, with the smearing width of $\gamma = 2$ MeV in 
the present calculation, the peak splitting disappears.
The double-peak structures in deformed nuclei for ISGMR, IVGMR, ISGDR,
and IV-LEOR, observed in the present calculation are all
associated with the coupling among GRs with different multipolarity.

Calculations using several commonly used Skyrme functionals
in the nuclear EDF method all give a fairly good reproduction of
the experimental data,
not only for the GRs but also for the low-lying collective modes in the spherical 
and the well deformed nuclei. 
Comparison of the GR results with
the experimental data obtained 
at RCNP~\cite{ito03} and TAMU~\cite{you04}
was performed in details
for the spherical nucleus $^{144}$Sm and the deformed $^{154}$Sm.
The experimental data for the ISGMR and the ISGDR 
indicates the incompressibility around $210-230$ MeV.
The excitation energy of the ISGQR is well reproduced with 
the effective mass $m^*_0/m \simeq 0.8 - 0.9$ both in $^{144}$Sm and in $^{154}$Sm.
The experimental data for the HEOR are very different between
the two experiments \cite{ito03,you04}.
A further experiments for HEOR are
needed to confirm the value of the effective mass.

\begin{acknowledgments}   
Valuable discussions with G.~Col\`o are acknowledged. 
The work is supported in part by Grant-in-Aid for Scientific Research
(Nos. 21340073, 20105003 and 23740223) 
and by the Joint Research Program at
Center for Computational Sciences, University of Tsukuba.
The numerical calculations were performed 
on RIKEN Integrated Cluster of Clusters (RICC), 
T2K at University of Tsukuba and SR16000 at the Yukawa Institute of Theoretical Physics, Kyoto University. 
\end{acknowledgments}

\begin{appendix}
\section{QRPA matrix elements}
Using the quasiparticle wave functions 
$\varphi_{1}(\boldsymbol{r}\sigma)$ and $\varphi_{2}(\boldsymbol{r}\sigma)$, 
the solutions of the coordinate-space HFB equation, 
the matrix elements appearing in the QRPA matrix are written as
\begin{align}
&A_{\alpha\beta\gamma\delta}=(E_{\alpha}+E_{\beta})\delta_{\alpha\gamma}\delta_{\beta\delta} \notag \\
&+\sum_{\sigma_{1},\sigma_{2},\sigma_{1}^{\prime},\sigma_{2}^{\prime}}
\int d\boldsymbol{r}_{1}d\boldsymbol{r}_{2} d\boldsymbol{r}_{1}^{\prime}d\boldsymbol{r}_{2}^{\prime} \times \notag \\
& \{ 
\varphi_{1,\alpha}(\boldsymbol{r}_{1}\bar{\sigma_{1}})
\varphi_{1,\beta}(\boldsymbol{r}_{2}\bar{\sigma_{2}})
\bar{v}_{pp}(12;1^{\prime}2^{\prime})
\varphi^{*}_{1,\gamma}(\boldsymbol{r}_{1}^{\prime}\bar{\sigma_{1}^{\prime}})
\varphi^{*}_{1,\delta}(\boldsymbol{r}_{2}^{\prime}\bar{\sigma_{2}^{\prime}}) \notag \\
&+
\varphi_{2,\alpha}(\boldsymbol{r}_{1}\sigma_{1})\varphi_{2,\beta}(\boldsymbol{r}_{2}\sigma_{2})
\bar{v}_{pp}(12;1^{\prime}2^{\prime})
\varphi^{*}_{2,\gamma}(\boldsymbol{r}_{1}^{\prime}\sigma_{1}^{\prime})
\varphi^{*}_{2,\delta}(\boldsymbol{r}_{2}^{\prime}\sigma_{2}^{\prime}) \notag \\
&-
\varphi_{1,\alpha}(\boldsymbol{r}_{1}\bar{\sigma_{1}})
\varphi^{*}_{2,\gamma}(\boldsymbol{r}_{2}\sigma_{2})
\bar{v}_{ph}(12;1^{\prime}2^{\prime})
\varphi_{2,\beta}(\boldsymbol{r}_{1}^{\prime}\sigma_{1}^{\prime})
\varphi^{*}_{1,\delta}(\boldsymbol{r}_{2}^{\prime}\bar{\sigma_{2}^{\prime}}) \notag \\
&-
\varphi_{1,\beta}(\boldsymbol{r}_{1}\bar{\sigma_{1}})
\varphi^{*}_{2,\delta}(\boldsymbol{r}_{2}\sigma_{2}) 
\bar{v}_{ph}(12;1^{\prime}2^{\prime})
\varphi_{2,\alpha}(\boldsymbol{r}_{1}^{\prime}\sigma_{1}^{\prime})
\varphi^{*}_{1,\gamma}(\boldsymbol{r}_{2}^{\prime}\bar{\sigma_{2}^{\prime}})
\notag \\
&+
\varphi_{1,\alpha}(\boldsymbol{r}_{1}\bar{\sigma_{1}})
\varphi^{*}_{2,\delta}(\boldsymbol{r}_{2}\sigma_{2})
\bar{v}_{ph}(12;1^{\prime}2^{\prime})
\varphi_{2,\beta}(\boldsymbol{r}_{1}^{\prime}\sigma_{1}^{\prime})
\varphi^{*}_{1,\gamma}(\boldsymbol{r}_{2}^{\prime}\bar{\sigma_{2}^{\prime}}) \notag \\
&+
\varphi_{1,\beta}(\boldsymbol{r}_{1}\bar{\sigma_{1}})
\varphi^{*}_{2,\gamma}(\boldsymbol{r}_{2}\sigma_{2})
\bar{v}_{ph}(12;1^{\prime}2^{\prime})
\varphi_{2,\alpha}(\boldsymbol{r}_{1}^{\prime}\sigma_{1}^{\prime})
\varphi^{*}_{1,\delta}(\boldsymbol{r}_{2}^{\prime}\bar{\sigma_{2}^{\prime}})
\}, \label{A_matrix_QRPA} \\
&B_{\alpha\beta\gamma\delta}=
\sum_{\sigma_{1},\sigma_{2},\sigma_{1}^{\prime},\sigma_{2}^{\prime}}
\int d\boldsymbol{r}_{1}d\boldsymbol{r}_{2} d\boldsymbol{r}_{1}^{\prime}d\boldsymbol{r}_{2}^{\prime}\times \notag \\
&\{-
\varphi_{1,\alpha}(\boldsymbol{r}_{1}{\bar \sigma_{1}})
\varphi_{1,\beta}(\boldsymbol{r}_{2}{\bar \sigma_{2}})
\bar{v}_{pp}(12;1^{\prime}2^{\prime})
\varphi_{2,\bar{\gamma}}(\boldsymbol{r}_{1}^{\prime}\sigma_{1}^{\prime})
\varphi_{2,\bar{\delta}}(\boldsymbol{r}_{2}^{\prime}\sigma_{2}^{\prime}) \notag \\
&-
\varphi_{2,\alpha}(\boldsymbol{r}_{1}\sigma_{1})
\varphi_{2,\beta}(\boldsymbol{r}_{2}\sigma_{2})
\bar{v}_{pp}(12;1^{\prime}2^{\prime})
\varphi_{1,\bar{\gamma}}(\boldsymbol{r}_{1}^{\prime}{\bar \sigma_{1}^{\prime}})
\varphi_{1,\bar{\delta}}(\boldsymbol{r}_{2}^{\prime}{\bar \sigma_{2}^{\prime}}) \notag \\
&+
\varphi_{1,\alpha}(\boldsymbol{r}_{1}{\bar \sigma_{1}})
\varphi_{1,\bar{\gamma}}(\boldsymbol{r}_{2}{\bar \sigma_{2}})
\bar{v}_{ph}(12;1^{\prime}2^{\prime})
\varphi_{2,\beta}(\boldsymbol{r}_{1}^{\prime}\sigma_{1}^{\prime})
\varphi_{2,\bar{\delta}}(\boldsymbol{r}_{2}^{\prime}\sigma_{2}^{\prime}) \notag \\
&+
\varphi_{1,\beta}(\boldsymbol{r}_{1}{\bar \sigma_{1}})
\varphi_{1,\bar{\delta}}(\boldsymbol{r}_{2}{\bar \sigma_{2}})
\bar{v}_{ph}(12;1^{\prime}2^{\prime})
\varphi_{2,\alpha}(\boldsymbol{r}_{1}^{\prime}\sigma_{1}^{\prime})
\varphi_{2,\bar{\gamma}}(\boldsymbol{r}_{2}^{\prime}\sigma_{2}^{\prime}) \notag \\
&-
\varphi_{1,\alpha}(\boldsymbol{r}_{1}{\bar \sigma_{1}})
\varphi_{1,\bar{\delta}}(\boldsymbol{r}_{2}{\bar \sigma_{2}})
\bar{v}_{ph}(12;1^{\prime}2^{\prime})
\varphi_{2,\beta}(\boldsymbol{r}_{1}^{\prime}\sigma_{1}^{\prime})
\varphi_{2,\bar{\gamma}}(\boldsymbol{r}_{2}^{\prime}\sigma_{2}^{\prime}) \notag \\
&-
\varphi_{1,\beta}(\boldsymbol{r}_{1}{\bar \sigma_{1}})
\varphi_{1,\bar{\gamma}}(\boldsymbol{r}_{2}{\bar \sigma_{2}})
\bar{v}_{ph}(12;1^{\prime}2^{\prime})
\varphi_{2,\alpha}(\boldsymbol{r}_{1}^{\prime}\sigma_{1}^{\prime})
\varphi_{2,\bar{\delta}}(\boldsymbol{r}_{2}^{\prime}\sigma_{2}^{\prime})
\}.
\end{align}
Here, the time-reversed state is defined as
\begin{equation}
\varphi_{\bar{i}}(\boldsymbol{r}\sigma)=-2\sigma \varphi_{i}^{*}(\boldsymbol{r} -\sigma).
\end{equation}

If one assumes that the effective pairing interaction is local, 
$\bar{v}_{pp}$ is written as
\begin{align}
\bar{v}_{pp}(12;1^{\prime}2^{\prime})=&
V_{pp}(\boldsymbol{r}_{1}\sigma_{1} \tau_1,\boldsymbol{r}_{2}\sigma_{2} \tau_2)\times \notag \\
& \delta(\boldsymbol{r}_{1}^{\prime}-\boldsymbol{r}_{1})
\delta_{\sigma_{1}^{\prime},\sigma_{1}}\delta_{\tau_1^{\prime},\tau_1}
\delta(\boldsymbol{r}_{2}^{\prime}-\boldsymbol{r}_{2})
\delta_{\sigma_{2}^{\prime},\sigma_{2}} \delta_{\tau_2^{\prime},\tau_2},
\end{align}
and for $V_{pp}$ we use the form 
\begin{align}
V_{pp}(\boldsymbol{r}_{1}\sigma_{1}\tau_1,\boldsymbol{r}_{2}\sigma_{2} \tau_2)= &
V_{0}\mathrm{g}_{q}[\varrho(\boldsymbol{r}_{1}),\varrho_{1}(\boldsymbol{r}_{1})] \times \notag \\
& 
\delta(\boldsymbol{r}_{1}-\boldsymbol{r}_{2})\delta_{\sigma_{1},-\sigma_{2}} \delta_{\tau_1,\tau_2}.
\end{align}
in the present paper. 

Similarly, the effective interaction for the p-h channel reads
\begin{align}
\bar{v}_{ph}(12;1^{\prime}2^{\prime})=&
V_{ph}(\boldsymbol{r}_{1}\sigma_{1}\tau_1,\boldsymbol{r}_{2}\sigma_{2}\tau_2)\times \notag \\
& \delta(\boldsymbol{r}_{1}^{\prime}-\boldsymbol{r}_{1})
\delta_{\sigma_{1}^{\prime},\sigma_{1}}\delta_{\tau_1^{\prime},\tau_1}
\delta(\boldsymbol{r}_{2}^{\prime}-\boldsymbol{r}_{2})
\delta_{\sigma_{2}^{\prime},\sigma_{2}}
\delta_{\tau_2^{\prime},\tau_2},
\end{align}
and we take the form
\begin{align}
&V_{ph}(\boldsymbol{r}_{1}\sigma_{1}\tau_1,\boldsymbol{r}_{2}\sigma_{2}\tau_2)= \notag \\
&(a_{0}+a_{0}^{\prime}\boldsymbol{\tau}_{1}\cdot\boldsymbol{\tau}_{2}+
(b_{0}+b_{0}^{\prime}\boldsymbol{\tau}_{1}\cdot\boldsymbol{\tau}_{2})
\boldsymbol{\sigma}_{1}\cdot\boldsymbol{\sigma}_{2})
\delta(\boldsymbol{r}_{1}-\boldsymbol{r}_{2}) \notag \\
&+(a_{1}+a_{1}^{\prime}\boldsymbol{\tau}_{1}\cdot\boldsymbol{\tau}_{2}+
(b_{1}+b_{1}^{\prime}\boldsymbol{\tau}_{1}\cdot\boldsymbol{\tau}_{2})
\boldsymbol{\sigma}_{1}\cdot\boldsymbol{\sigma}_{2}) \notag \\
& \times (\boldsymbol{k}^{\dagger 2}\delta(\boldsymbol{r}_{1}-\boldsymbol{r}_{2})
+ \delta(\boldsymbol{r}_{1}-\boldsymbol{r}_{2}) \boldsymbol{k}^{2}) \notag \\
& + (a_{2}+a_{2}^{\prime}\boldsymbol{\tau}_{1}\cdot\boldsymbol{\tau}_{2}+
(b_{2}+b_{2}^{\prime}\boldsymbol{\tau}_{1}\cdot\boldsymbol{\tau}_{2})
\boldsymbol{\sigma}_{1}\cdot\boldsymbol{\sigma}_{2}) \notag \\
& \times (\boldsymbol{k}^{\dagger}\cdot\delta(\boldsymbol{r}_{1}-\boldsymbol{r}_{2})\boldsymbol{k} ) \notag \\
&+ (a_{4}+a_{4}^{\prime}\boldsymbol{\tau}_{1}\cdot\boldsymbol{\tau}_{2})
(\boldsymbol{\sigma}_{1}+\boldsymbol{\sigma}_{2})\cdot
\boldsymbol{k}^{\dagger}\times \delta(\boldsymbol{r}_{1}-\boldsymbol{r}_{2})\boldsymbol{k}
\label{v_res_ph}
\end{align}
with the standard notations of $\boldsymbol{k}$ and $\boldsymbol{k}^{\dagger}$. 
The coefficients in Eq.~(\ref{v_res_ph}) are given in Ref.~\cite{ter05}. 
The coefficients $a_{0}, a_{0}^{\prime}, b_{0}$ and $b_{0}^{\prime}$ 
are density dependent and include the rearrangement terms. 
In the present paper, 
we have an additional contribution to these terms coming from the pairing EDF~(\ref{pair_int}). 
They are
\begin{eqnarray}
\left \{
\begin{array}{lll}
-\dfrac{V_{0}}{2}\dfrac{\eta_{2}}{\varrho_{0}^{2}} 
[\tilde{\varrho}_{\nu}^{2}(\boldsymbol{r})+\tilde{\varrho}_{\pi}^{2}(\boldsymbol{r})] &
(\mathrm{for} & \nu-\nu, \pi-\pi) \\
\dfrac{V_{0}}{2}\dfrac{\eta_{2}}{\varrho_{0}^{2}} 
[\tilde{\varrho}_{\nu}^{2}(\boldsymbol{r})+\tilde{\varrho}_{\pi}^{2}(\boldsymbol{r})] &
(\mathrm{for}&  \nu-\pi).
\end{array}
\right.
\end{eqnarray}

\section{Parameters of the giant resonances}
We summarize here the peak energy and the width of the 
GRs obtained by the calculations with the SkM* functional.

\begin{table*}[b]
\begin{center}
\caption{The parameters of the ISGMR and IVGMR. }
\label{GMR}
\begin{tabular}{ccccccccc}
\hline \hline
\noalign{\smallskip}
 &  \multicolumn{4}{c}{ISGMR}   & \multicolumn{4}{c}{IVGMR}  \\
 & $E_{x}$ & $\Gamma$  & $E_{x}$ & $\Gamma$ 
 & $E_{x}$ & $\Gamma$  & $E_{x}$ & $\Gamma$ \\
 & (MeV)  & (MeV)  & (MeV) & (MeV) 
 & (MeV)  & (MeV)  & (MeV) & (MeV) \\
\noalign{\smallskip}\hline\noalign{\smallskip}
$^{142}$Nd  &  & & 15.0  & 2.67 & & & 30.0 & 10.7   \\
$^{144}$Nd  &  & & 14.5 & 2.79 & & & 29.6 & 10.2   \\
$^{146}$Nd  & 12.1 & 2.37 & 14.8 & 3.05 & 21.9 & 7.47 & 29.7 & 9.68 \\
$^{148}$Nd  & 11.9 & 2.83 & 15.0 & 3.05 & 21.7 & 4.54 & 29.8 & 9.39 \\
$^{150}$Nd  & 11.8 & 3.22 & 15.6 & 3.15 & 21.1 & 3.92 & 30.2 & 9.81  \\
$^{152}$Nd  & 11.5 & 3.40 & 15.7 & 3.20 & 20.7 & 3.91 & 30.3 & 9.76 \\
$^{144}$Sm  &  & & 14.9 & 2.62  & & & 29.9 & 10.9   \\
$^{146}$Sm  &  & & 14.4 & 2.68  & & & 29.4 & 10.4   \\
$^{148}$Sm  & 12.2 & 2.07 & 14.7 & 2.97 & 21.4 & 6.28 & 29.5 & 10.0 \\
$^{150}$Sm  & 11.9 & 2.79 & 15.0 & 2.97 & 21.6 & 4.27 & 29.8 & 9.52 \\
$^{152}$Sm  & 11.8 & 3.20 & 15.5 & 3.04 & 21.2 & 3.79 & 30.2 & 9.90 \\
$^{154}$Sm  & 11.5 & 3.39 & 15.6 & 3.12 & 20.9 & 3.77 & 30.3 & 9.80 \\
\noalign{\smallskip}
\hline \hline
\end{tabular}
\end{center}
\end{table*}

\begin{table*}[t]
\begin{center}
\caption{The parameters of the ISGDR and IVGDR.}
\label{GDR}
\begin{tabular}{ccccccccccc}
\hline \hline
\noalign{\smallskip}
 & \multicolumn{2}{c}{LE-ISGDR} & \multicolumn{4}{c}{ISGDR}  
 & \multicolumn{4}{c}{IVGDR}  \\
 & $E_{x}$ & $\Gamma$
 & $E_{x}$ & $\Gamma$  & $E_{x}$ & $\Gamma$ 
 & $E_{x}$ & $\Gamma$  & $E_{x}$ & $\Gamma$ \\
 & (MeV)  & (MeV)  & (MeV) & (MeV) 
 & (MeV)  & (MeV)  & (MeV) & (MeV)
 & (MeV) & (MeV) \\
\noalign{\smallskip}\hline\noalign{\smallskip}
$^{142}$Nd  & 14.2 & 7.62 & 26.0 & 6.32 &  &  & 14.8 & 4.40 & &   \\
$^{144}$Nd  & 13.9 & 8.25 & 25.9 & 6.33 & & & 14.8 & 4.34 & &    \\
$^{146}$Nd  & 13.8 & 8.90 & 23.4 & 7.49 & 26.7 & 5.15 & 14.1& 3.65 & 17.0 & 3.20  \\
$^{148}$Nd  & 13.8 & 9.26 & 22.3 & 5.78 & 26.7 & 5.71 & 13.5 & 3.58 & 16.5 & 4.73  \\
$^{150}$Nd  & 13.7 & 11.3 & 21.7 & 7.62 & 27.1 & 5.74 & 12.4 & 2.56  & 15.7 & 5.65   \\
$^{152}$Nd  & 13.6 & 14.1 & 21.1 & 9.26 & 27.2 & 6.63 & 12.0 & 2.56  & 15.7 & 5.69   \\
$^{144}$Sm & 14.3 & 9.52 & 25.9 & 6.20 &  &  & 14.8 & 4.38  &  &   \\
$^{146}$Sm & 13.9 & 10.5 & 25.8 & 6.21 &  & & 14.8 & 4.31 & &  \\
$^{148}$Sm & 14.0 & 10.3 & 23.6 & 7.59 & 26.7 & 4.95 & 14.1 & 3.58 & 16.9 & 3.45 \\
$^{150}$Sm & 14.0 & 9.77 & 22.2 & 5.69 & 26.6 & 5.87 & 13.3 & 3.30 & 16.0 & 4.96 \\
$^{152}$Sm & 14.0 & 10.8 & 21.4 & 6.44 & 26.8 & 7.87 & 12.4 & 2.46  & 15.7 & 5.68   \\
$^{154}$Sm & 14.0 & 12.6 & 21.0  & 8.21 & 26.9 & 7.44  & 12.1 & 2.51 & 15.7 & 5.70   \\
\noalign{\smallskip}
\hline \hline
\end{tabular}
\end{center}
\end{table*}

\begin{table}[t]
\begin{center}
\caption{The parameters of the ISGQR and IVGQR. }
\label{GQR}
\begin{tabular}{ccccc}
\hline \hline
\noalign{\smallskip}
&  \multicolumn{2}{c}{ISGQR}   & \multicolumn{2}{c}{IVGQR}  \\
& $E_{x}$ & $\Gamma$  & $E_{x}$ & $\Gamma$  \\
& (MeV)  & (MeV)  & (MeV) & (MeV)  \\
\noalign{\smallskip}\hline\noalign{\smallskip}
$^{142}$Nd  & 13.3 & 2.89  & 24.8 & 5.20 \\
$^{144}$Nd  &12.9  & 2.93 & 24.5 & 5.12 \\
$^{146}$Nd  & 12.7 & 3.01 & 24.0 & 5.71 \\
$^{148}$Nd  & 12.6 & 3.51 & 23.5 & 6.69 \\
$^{150}$Nd  & 12.7 & 4.71 & 23.7 & 8.42 \\
$^{152}$Nd  & 12.5 & 5.23 & 23.5 & 9.11 \\
$^{144}$Sm  & 13.3 & 2.73 & 24.8 & 4.96 \\
$^{146}$Sm  & 12.9 & 2.77 & 24.5 & 4.91 \\
$^{148}$Sm & 12.7 & 3.02 & 24.2 & 5.59 \\
$^{150}$Sm  & 12.6 & 3.63 & 23.8 & 6.66  \\
$^{152}$Sm  & 12.7 & 4.71  & 23.7 & 8.06  \\
$^{154}$Sm  & 12.6 & 5.14  & 23.5 & 8.64  \\
\noalign{\smallskip}
\hline \hline
\end{tabular}
\end{center}
\end{table}

\begin{table*}[t]
\begin{center}
\caption{The parameters of the HEOR and IVGOR.}
\label{GOR}
\begin{tabular}{ccccccccc}
\hline \hline
\noalign{\smallskip}
 &  \multicolumn{2}{c}{HEOR}   & \multicolumn{4}{c}{IV-LEOR} 
 & \multicolumn{2}{c}{IV-HEOR}  \\
 & $E_{x}$ & $\Gamma$  & $E_{x}$ & $\Gamma$ 
 & $E_{x}$ & $\Gamma$  & $E_{x}$ & $\Gamma$ \\
 & (MeV)  & (MeV)  & (MeV) & (MeV)  
 & (MeV)  & (MeV)  & (MeV) & (MeV) \\
\noalign{\smallskip}\hline\noalign{\smallskip}
$^{142}$Nd & 24.1 & 3.65 & 12.5 & 6.97 & & & 33.3 & 8.02  \\
$^{144}$Nd & 24.0 & 3.73 & 12.4 & 7.68 & & & 33.1 & 7.85   \\
$^{146}$Nd & 23.8 & 4.44 & 12.2 & 9.94 & & & 32.8 & 8.01 \\
$^{148}$Nd & 23.5 & 5.31 & 11.8 & 8.26 & 16.4 & 4.76 & 32.4 & 8.38 \\
$^{150}$Nd & 23.2 & 6.47 & 11.5 & 6.29 & 16.0 & 4.86 & 32.0 & 9.28  \\
$^{152}$Nd & 22.9 & 6.84 & 11.5 & 6.01 & 16.0 & 4.55 & 31.7 & 9.73  \\
$^{144}$Sm & 24.0 & 3.70 & 12.4 & 6.83 & & & 33.2 & 7.78  \\
$^{146}$Sm & 24.0 & 3.66 & 12.3 & 7.48 & & & 33.1 & 7.61   \\
$^{148}$Sm & 23.8 & 4.41 & 12.2 & 9.63 & & & 32.7 & 7.83   \\
$^{150}$Sm & 23.4 & 5.51 & 11.9 & 8.28 & 16.2 & 4.56 & 32.3 & 8.32   \\
$^{152}$Sm & 23.1 & 6.84 & 11.8 & 6.36 & 16.1 & 4.49 & 32.0 & 9.00  \\
$^{154}$Sm & 22.9 & 6.74 & 11.7 & 6.10 & 16.1 & 4.21 & 31.7 & 9.34  \\
\noalign{\smallskip}
\hline \hline
\end{tabular}
\end{center}
\end{table*}

\end{appendix}

\end{document}